\documentclass[acmsmall]{acmart}

\usepackage{array}
\usepackage[fixed]{fontawesome5}
\faStyle{regular}
\usepackage{xcolor}
\usepackage{subcaption}
\usepackage{mdframed}
\usepackage{booktabs}
\usepackage{caption}

\AtBeginDocument{%
  \providecommand\BibTeX{{%
    \normalfont B\kern-0.5em{\scshape i\kern-0.25em b}\kern-0.8em\TeX}}}

\renewcommand\footnotetextcopyrightpermission[1]{} 
\makeatletter
\renewcommand\@formatdoi[1]{\ignorespaces}
\def\@eprintstripzeros#1{}
\makeatother

\begin{document}



\title[LLM Agents for Improving Engagement with Interventions]{Large Language Model Agents for Improving Engagement with Behavior Change Interventions: Application to Digital Mindfulness}

\author{Harsh Kumar}
\email{harsh@cs.toronto.edu}
\orcid{0000-0003-2878-3986}
\affiliation{%
  \institution{Computer Science, University of Toronto}
  \country{Canada}
}
\author{Suhyeon Yoo}
\email{suhyeon.yoo@mail.utoronto.ca}
\affiliation{%
  \institution{Computer Science, University of Toronto}
  \country{Canada}
}
\author{Angela Zavaleta Bernuy}
\email{angela.zavaletabernuy@mail.utoronto.ca}
\affiliation{%
  \institution{Computer Science, University of Toronto}
  \country{Canada}
}
\author{Jiakai Shi}
\email{jiakai.shi@mail.utoronto.ca}
\affiliation{%
  \institution{Computer Science, University of Toronto}
  \country{Canada}
}
\author{Huayin Luo}
\email{huayin.luo@mail.utoronto.ca}
\affiliation{%
  \institution{Computer Science, University of Toronto}
  \country{Canada}
}
\author{Joseph Williams}
\email{williams@cs.toronto.edu}
\affiliation{%
  \institution{Computer Science, University of Toronto}
  \country{Canada}
}
\author{Anastasia Kuzminykh}
\email{anastasia.kuzminykh@utoronto.ca}
\affiliation{%
  \institution{School of Information, University of Toronto}
  \country{Canada}
}
\author{Ashton Anderson}
\email{ashton@cs.toronto.edu}
\affiliation{%
  \institution{Computer Science, University of Toronto}
  \country{Canada}
}
\author{Rachel Kornfield}
\affiliation{%
  \institution{Preventive Medicine, Northwestern University}
  \city{Chicago}
  \state{Illinois}
  \country{USA}
}

\renewcommand{\shortauthors}{Harsh Kumar et al.}

\begin{abstract}



Although engagement in self-directed wellness exercises typically declines over time, integrating social support such as coaching can sustain it. However, traditional forms of support are often inaccessible due to the high costs and complex coordination. Large Language Models (LLMs) show promise in providing human-like dialogues that could emulate social support. Yet, in-depth, in situ investigations of LLMs to support behavior change remain underexplored. We conducted two randomized experiments to assess the impact of LLM agents on user engagement with mindfulness exercises. First, a single-session study, involved 502 crowdworkers; second, a three-week study, included 54 participants. We explored two types of LLM agents: one providing information and another facilitating self-reflection. Both agents enhanced users' intentions to practice mindfulness. However, only the information-providing LLM, featuring a friendly persona, significantly improved engagement with the exercises. Our findings suggest that specific LLM agents may bridge the social support gap in digital health interventions.

\end{abstract}




\maketitle

\section{Introduction}


Changing behavior to improve wellness is challenging, even when the potential benefits are obvious \cite{milkman2021megastudies, bouton2014behavior, middleton2013long}. 
For example, people are often motivated to participate in traditional yoga classes, try meditation routines discovered in books or magazines, participate in community-led mindfulness sessions, or explore wellness exercises on digital platforms such as Instagram and YouTube. However, forming habits and following through with these exercises proves to be a significant hurdle \cite{yardley2016understanding, monge2019race}. This lack of perseverance can be attributed to several factors, including lack of accountability and the struggle to overcome ambivalence \cite{bouton2014behavior, kelly2016changing, kelly1991prediction}. Nevertheless, there is evidence that social support can contribute to the success of behavior change initiatives by increasing accountability, motivation, and emotional sustenance \cite{o2018suddenly, boothroyd2010peers, alrobai2018building, moore2019effectiveness, lee2015understanding}.

The `Supportive Accountability' model proposed by Mohr \textit{et al.} \cite{mohr2011supportive} suggests that digital health promotion tools are more likely to be effective when complemented by human support. This support goes beyond simple companionship, incorporating benevolence, relevant expertise, and tailored guidance. An example could be a \textit{coach} that provides information, scaffolding the learning process with factual and procedural knowledge \cite{schueller2015initial, mohr2011supportive}. On the other hand, there can be support that operates through inquiry and reflection, which encourages users to engage in self-exploration and critical thinking about their behaviors and underlying motivations \cite{kocielnik2018designing, tian2021let}. Integrating social and expertise-based support into technology may facilitate long-term behavioral change.
However, traditional forms of support are inaccessible to many due to the costs and complexities of providing support \cite{agapie2016plansourcing}. Human support is often the most expensive and complex part of a digital health intervention \cite{mohr2021banbury, lattie2022overview, barras2019mental}. Some people also have preferences not to interact directly with others or lack existing supportive relationships that can be leveraged to support new practices.

Recently, advances in deep learning have led to the development of LLMs such as GPT-4 \cite{brown2020language, achiam2023gpt}. These models are large neural networks trained on vast collections of textual data, enabling them to process inputs in natural language and generate human-like text. Conversations, essential for providing social support among humans, can now be convincingly mimicked by LLM-based conversational agents, simulating the personalized and open-ended dimensions of human-to-human interactions that, until now, have been challenging to achieve in automated conversation systems \cite{shanahan2023role, meyerhoff2024small, stade2024large}. 
LLMs have potential to play a multifaceted role in helping individuals achieve behavior change. Although significant research has explored social support in various contexts, less is known about the effectiveness of LLM-based agents for long-term behavior change in ecologically valid settings \cite{stade2024large}.

Our research draws from existing CSCW and social support literature to understand the role of LLM-based agents in providing social support to adopt wellness practices \cite{agapie2016plansourcing, lee2015understanding, miller2012work}. Specifically, we focus on evaluating the impact of LLM agents on two key aspects: the users' intention to participate in wellness exercises and their actual engagement with these exercises. We also gather user perspectives on the support provided by LLM agents to understand how people think about collaborating with LLM agents around behavior change. We address the following research questions.
\vspace*{0.5em}
\begin{description}
    \hrule height 0.1em
    \item[RQ1] (Intention to Practice Exercises): How do LLM agents influence users’ intention to engage with wellness exercises?
    \item[RQ2] (Engagement with Exercises): What impact do LLM agents have on the actual engagement with wellness exercises, as indicated by the initiation and completion rates of exercises?
    \item[RQ3] (Measures of Well-being): What impact do LLM agents have on immediate and long-term well-being measures (such as stress and mindfulness scores) related to wellness exercises?
\end{description}
\hrule height 0.025em
\vspace*{0.5em}


We focus on mindfulness as a case study, a practice increasingly recognized for its extensive benefits in improving mental health, reducing stress, improving emotional regulation, and fostering overall well-being \cite{niksirat2019attention, terzimehic2019review, li2022beyond, laurie2016making}. 
A popular method of learning mindfulness is through tutorial videos.
This approach allows people to learn about mindfulness and practice it at their own pace and convenience. However, despite the accessibility and flexibility of these self-guided mindfulness programs, they often face challenges with attrition. Studies have shown that consistent engagement in such self-directed mindfulness interventions can be challenging, and many participants do not complete the programs \cite{ schwartz2023mindfulness, winter2022engagement, baumel2019objective}. Estimates suggest that 23-39\% people drop out within 10 weeks of the program \cite{linardon2023rates}.
There is an opportunity to enhance engagement with self-directed wellness practices through additional social support. The nature of this social support can vary to meet individual needs at any point in the behavior change process \cite{meyerhoff2022meeting}. Informational support and validation can play crucial roles in helping people understand the benefits of change and motivate them to begin practicing. This type of support can provide the necessary clarity and encouragement that individuals might need at various times during their wellness journey. In addition, planning support and decisional reinforcement can help individuals stay on track and effectively integrate practices into their daily routines. Self-reflection can allow participants to introspect and evaluate their progress \cite{ploderer2014social, li2011understanding, gao2012design}. Reflective support is particularly valuable as it can deepen the understanding of personal experiences and promote sustained behavior change \cite{baumer2014reviewing, baumer2015reflective, bentvelzen2022revisiting}. In this paper, we use LLM agents to provide informational and reflective support to investigate whether these agents can meet varying needs for support throughout the mindfulness journey of an individual.

We conducted a formative study involving online crowdworkers ($N=502$) to capture user perspectives and intentions about mindfulness practice when supplemented with different forms of LLM agent support (RQ1). As design probes, we developed two types of LLM agents using GPT-3: the Information Chatbot, designed to provide detailed information about mindfulness practices, and the Reflection Chatbot, designed to help participants reflect on their mindfulness experiences and plan their future mindfulness sessions. 
Additionally, our study incorporated a tutorial video as a fundamental element of many digital interventions promoting mindfulness and similar wellness behaviors. Overall, the design of the study was a 2 (\textit{Information Chatbot}: present vs. absent) x 2 (\textit{Tutorial Video}: present vs. absent) x 2 (\textit{Reflection Chatbot}: present vs. absent) factorial between-subjects experiment. 
Although both chatbots were perceived to be useful, user feedback suggested a preference for more relaxed and friendly conversational interactions (small talk), which they found particularly lacking in the Information Chatbot. Self-reported intentions to practice mindfulness were higher among participants who first interacted with the Information Chatbot before watching the video and those who engaged with the Reflection Chatbot after the video, compared to the control group that did not use the chatbots. Participants mentioned repetitiveness in the responses as a point of discontent for both chatbots.

However, as highlighted in previous research \cite{sheeran2016intention, rhodes2013big}, such preliminary insights require validation through real-world deployments to gauge the alignment between expressed intentions and actual user behavior. To this end, we conducted a 3-week deployment study with 54 participants interested in learning about mindfulness (RQ2). The core intervention included instructional videos and email reminders every two days to practice mindfulness. In a 2 (\textit{Sociable Information LLM agent}: present vs. absent) x 2 (\textit{Reflection LLM agent post Exercise}: present vs. absent) between-subjects experiment, participants were randomized to have access to the two LLM chatbots. Based on the participants' feedback in the formative study, the Information Chatbot was developed into a sociable informational companion chatbot, \textit{Mindy}, characterized by a friendly persona. Half of the participants were randomly assigned to receive access to \textit{Mindy} in their reminder emails. We continued using the Reflection LLM agent, which was enhanced based on initial feedback and used immediately after each instructional video. Again, half of the participants were randomly given access to this reflective component after each session. We upgraded to GPT-4 from GPT-3 to address issues like repetitiveness mentioned by participants in the formative study. Participants who had access to the Sociable Information LLM, \textit{Mindy}, showed a notable increase in engagement with a 18\% higher initiation rate and a 12\% higher completion rate of daily exercises compared to those without access to the agent (\(p < 0.01\)). Interestingly, incorporating the Reflection LLM agent following video exercises did not significantly affect initiation or completion rates. This finding was somewhat unexpected, as the \textit{Reflection LLM}, with its specific objective of aiding in planning and reflection, was expected to play a more active role in enhancing exercise adherence.

These findings suggest that LLM agents can potentially improve people's intentions to participate in self-guided wellness exercises and their engagement with them. However, the findings also highlight that these LLM agents should be tested in a longitudinal setting to gauge the actual benefits before deployment. Given the increasing need for social support in our society, our research underscores the potential benefits of using LLMs to offer the necessary support, thus contributing to improved well-being and healthier lifestyles.

\section{Related Work}
\label{section:background}
Our work builds on long-standing research on influencing engagement in behavior change interventions, the role of social support in maintaining this engagement, technology-enabled mindfulness, conversational agents, and the growing literature on the interaction between humans and LLMs.

\subsection{User Engagement in Behavior Change Interventions}
\label{section:background-engagement}
Ensuring sustained engagement with behavior change tools is crucial for their effectiveness \cite{yardley2016understanding, karapanos2015sustaining}. Most behavior change interventions, especially those delivered through the digital medium without human support, suffer from high dropout and attrition \cite{eysenbach2005law, kohl2013online, yardley2016understanding}. Engagement alone may not directly result in behavior change, but it serves as an essential precondition \cite{yardley2016understanding, karapanos2015sustaining, graham2019experimental}. In an era marked by fragmented attention, distractions, and shifting priorities, a key challenge is designing technologies that maintain user interest without contributing to the noise of daily life \cite{greenfield2017radical, thackara2006bubble, de2017computational}. Karapanos \cite{karapanos2015sustaining} proposed the use of `Creating checking habits' powered by instant information rewards, similar to those in social media updates and emails. Although there is a risk that these habits could lead to addictive behaviors \cite{karapanos2015sustaining, gouveia2016exploring}, they also promise to foster beneficial engagement, such as encouraging more frequent exercise through interactions with the app.

Gouveia \textit{et al.} \cite{gouveia2015we} explored how behavior-change applications might create checking habits by continuously updating their feedback, similar to strategies employed in the computer-gaming and airline industries. This continuous update could maintain the informational reward from  app checking, which is believed to be a primary driver behind the formation of checking habits. Furthermore, enhancing engagement through social interactions, which we cover in the following subsection, could be another strategy to sustain engagement. Erickson and Kellogg's \cite{erickson2000social} concept of ``Social Translucence'' suggests shifting the responsibility of behavior change to families and social ties, promoting awareness of each other's behaviors through technology, rather than relying solely on the technology's persuasive capabilities. This shift towards social awareness and nudging within personal relationships could lead to a more sustainable change in behavior \cite{mohr2011supportive, kellogg2002social, barreto2013initial}.

Behavior change is a broad concept that can capture changes in how individuals prevent or manage health conditions, and their day-to-day practices and habits that have implications for general well-being. Here, we focus on the latter, considering the specific context of in initiating and sustaining mindfulness practice.

\subsection{Social Support in Wellness Behavior Change Journeys}
\label{section:background-support}
Social support is essential in influencing health behaviors and outcomes, acting as a cornerstone of physical and mental well-being \cite{thoits2011mechanisms, lee2015understanding}. Studies, such as those conducted by Tay \textit{et al.} \cite{tay2013social}, have highlighted the multifaceted benefits of social relationships, highlighting their importance in chronic illness self-management and mitigating suicidal tendencies. The expanding research in this field has led to diverse conceptualizations and operational frameworks for social support, reflecting its complexity and the variety of forms it can take \cite{house1988structures, liu2017assessing}.

Barrera and Ainlay \cite{barrera1983structure} provided a comprehensive classification of social support into six categories, including: Material Aid, signifying the provision of tangible resources like money or goods; Behavioral Assistance, which involves sharing tasks or providing physical labor; Intimate Interaction, encompassing traditional counseling behaviors such as listening and expressing empathy; Guidance, offering advice or information; Feedback, providing insights into an individual's behavior, thoughts, or feelings; and Positive Social Interaction, facilitating leisure and relaxation through social activities. Some forms of support are particularly relevant to the context of supporting wellness exercises. Material aid and behavioral assistance are less relevant for our specific context of self-directed mindfulness exercises. Guidance and Feedback can be synthesized into a singular category as `Directive Guidance' \cite{hirsch1980natural}. 
In contrast, intimate and positive social interactions can be viewed as elements of `Non-directive Support' \cite{gottlieb1978development}. 

Positive social interaction may be especially critical in well-being interventions, offering fun and relaxation opportunities essential for a balanced lifestyle. Supportive peer relationships, as found by Roach \cite{roach2018supportive}, significantly improved adolescents' mental health, indicating the broader implications of positive social interactions across different demographics. Although several self-directed programs and apps are available in the particular case of mindfulness, a common challenge remains in the lack of sustained use of these tools or strategies in daily life \cite{linardon2023rates, winter2022engagement}. Mohr \textit{et al.} \cite{mohr2011supportive} proposed a model for ``Supportive Accountability'', highlighting the critical role of social support in enhancing adherence and engagement in behavior change programs. The model suggests that incorporating social support elements in technology-enabled services significantly increases their effectiveness in promoting long-term behavioral changes \cite{tang2017neuroscience, wilson2020mindfulness}. Recent CSCW research underscores the importance of interpersonal relationships in the management of mental health concerns and the crucial role of technology in providing ongoing support \cite{burgess2019think, murnane2018personal}.

However, the scalability of providing high-quality social support presents a major challenge, as traditional methods, such as individual, in-person instruction, are not feasible on a large scale. Therefore, there is a need to explore and implement scalable solutions that can provide high-quality social support effectively and sustainably to a broader audience. Our research looks into how LLM-based agents can effectively embody different forms of Directive and Non-directive support, one that is informational versus one that encourages self-reflection, for a nuanced understanding of the varying needs for social support during behavior change.

\subsection{Technology-Enabled Mindfulness}

The HCI and CSCW literature has extensively explored the role of technology in supporting meditation to practice mindfulness \cite{terzimehic2019review}. 
Technology serves as a valuable tool, allowing users to practice in their own time and space and aiding users in achieving and sustaining focus on inner or outer experiences, such as breathing \cite{paredes2018just, patibanda2017life}, tactile movements \cite{hussien2017understanding, salehzadeh2017framework}, or even monitoring brain activity \cite{kosunen2017neuroadaptive, kosunen2016relaworld, sas2015meditaid}.  These techniques encompass a spectrum of applications, including Virtual Reality (VR), wearables, and online interventions.
Wearable devices have been investigated for their role in mindfulness practice, with a focus on monitoring biometrics using integrated sensors. Research has explored sensor placement options like the thumb, head, or infraclavicular locations to determine the most sensitive accelerometer location for measuring body motion during short meditation sessions \cite{rodriguez2018wearable}. Additionally, MeditAid, a headset providing binaural feedback for enhancing meditative states \cite{sas2015meditaid}, and the MindfulWatch, which tracks real-time breathing patterns during meditation using smartwatch technology \cite{hao2017mindfulwatch}, have emerged. These wearables extend to wrist-worn devices that can distinguish meditation periods by analyzing heart rate variability \cite{alvarez2021meditation, chiovato2021evaluation}. In workplace settings, brain-sensing wearable devices like the MUSE-S™ have been utilized for guided meditation sessions to reduce stress \cite{ghosh2023mindfulness}. 

While wearable devices, including VR head-mounted units, show promise in various applications, their reach among diverse users in daily contexts, particularly those new to mindfulness seeking accessible entry points, remains limited. Conversely,
numerous online programs and applications have been developed to promote mindfulness practices, leveraging the convenience of smartphones, tablets, and desktop systems for personal use.
These apps offer various features, including guided audio meditations, daily mindfulness activities, and training programs. 
For instance, apps offering guided audio meditations were associated with reduced blood pressure and significant enhancements in overall well-being, distress reduction, and the alleviation of job-related stress \cite{bostock2019mindfulness}.  Additionally, smartphone-based interventions provide users with simple daily mindfulness activities, further contributing to improved mental well-being \cite{howells2016putting}.  Walsh et al. \cite{walsh2019effects} highlighted that mindfulness training through smartphone apps can enhance subjective well-being, attentional control, and interoceptive integration. An intervention using a mindfulness meditation app, Calm, was associated with improved sleep quality, reduced alcohol consumption (particularly binge drinking), increased physical activity, and healthier dietary choices, including increased fruit and vegetable consumption \cite{huberty2019efficacy}. However, these apps suffer from high attrition rates \cite{linardon2023rates}.


\subsection{Conversational Agents for Behavioral Interventions}
\label{section:background-ca}
Conversational agents, or chatbots, have been increasingly deployed in various domains to positively influence user behavior, from improving sleep habits \cite{rick2019sleepbot} and offering psychiatric counseling \cite{oh2017chatbot} to promoting healthier diets, exercise routines \cite{huang2018chatbot}, smoking cessation \cite{calvaresi2019social}, and other health-related behavior change. Lukoff \textit{et al.} \cite{lukoff2018tablechat} introduced a chat-based food journaling tool to help families achieve their healthy eating goals, exemplifying the utility of chatbots in nurturing healthier lifestyle choices. In addition, chatbots have been used in professional settings to help users reflect on experiences \cite{kocielnik2018designing} or manage psychological transitions related to work \cite{williams2018supporting}, ultimately strengthening recovery and overall well-being. Educational initiatives have also seen the integration of pedagogically focused chatbots designed to improve learning outcomes \cite{kumar2023impact}. For example, Zvereva \textit{et al.} \cite{zvereva2020method} proposed a dialogue-based approach to assess and improve student motivation.

The role of chatbots extends into the mental health domain, where they have been used to manage symptoms and improve quality of life \cite{abd2020effectiveness}. In particular, Woebot \cite{fitzpatrick2017delivering}, a chatbot based on Cognitive Behavioral Therapy (CBT), demonstrated effectiveness in a randomized controlled trial, highlighting the potential of chatbots to provide mental health support. In addition, chatbots have facilitated self-disclosure \cite{lee2020hear, lee2020designing} and promoted self-compassion \cite{lee2019caring}, crucial components of personal well-being and behavioral change interventions.

Our investigation is based on existing research surrounding the design and implementation of conversational agents for behavior change and well-being \cite{cameron2018best}. Using LLMs for chatbots presents promising prospects and novel challenges \cite{jo2023understanding}. LLMs' ability to generate adaptive human-like responses offers an exciting frontier for behavior change technologies. However, LLMs also introduce ethical and safety concerns that surpass those associated with traditional chatbots, necessitating rigorous evaluations to understand their implications fully \cite{gabriel2024ethics, xu2024mental, de2023benefits}. There are also questions regarding how users feel about social support delivered by an LLM. Through laboratory and field deployments, our work addresses these critical questions, exploring the effectiveness of LLM-based chatbots in real-world settings. In addition, we explore the applicability and desirability of different forms of LLM-provided support, contributing to the literature on the types of automated social support to which users are the most receptive \cite{meng2021emotional, brandtzaeg2022my, laestadius2022too, nass1994computers}.

\subsection{Situating Work in Broader CSCW \& HCI Literature}


As we saw in previous sections, the current literature emphasizes the pivotal role of human support in facilitating wellness practices, particularly in self-directed exercises (Section \ref{section:background-engagement} and \ref{section:background-support}). Within the CSCW community, there is a growing emphasis on understanding the needs and preferences of people who pursue self-directed wellness and the development of supportive technologies \cite{meyerhoff2022meeting, kornfield2020energy, lattie2020designing, miller2012work}. Human facilitators in various formats, from personal coaches available in popular mindfulness apps to instructors leading group sessions, have significantly improved engagement with interventions and outcomes \cite{mohr2011supportive, geerts2021brainstorming, lee2015understanding}. These facilitators offer personalized guidance, emotional support, and accountability, aspects crucial to the success of such interventions \cite{geerts2021brainstorming}. Despite their effectiveness, human-led support faces limitations in cost, scalability, and accessibility \cite{mohr2021banbury, lattie2022overview, barras2019mental, agapie2016plansourcing}. In contrast, LLM-based agents, which can be made available everywhere and at a lower cost, emerge as a promising alternative. These agents can provide continuous judgment-free social support without time and location constraints and maintain user anonymity and privacy, making them a practical solution for wide-scale implementations \cite{memmert2023towards, yang2024social}.

Trained on extensive datasets, LLMs have the potential to adapt their guidance to various user contexts and challenges, a capability that is still being explored and refined \cite{chaszczewicz2024multi}. They can also support people with social anxiety or who, for various reasons, are uncomfortable interacting with human coaches about their well-being \cite{maples2024loneliness}. LLMs can serve as informative guides \cite{krause2017critique}, help users acquire new skills by offering precise and detailed explanations that reflect the user's specific contexts and concerns \cite{zhang2023bootstrap}, and resolve uncertainties about effectively carrying out wellness practices \cite{ma2023understanding, xu2024mental}. They can provide proactive check-ins, inquire about progress, and assist in planning future wellness activities, thus promoting accountability. Furthermore, their ability to engage in empathetic and supportive dialogues positions them as a `friendly presence,' offering encouragement and reflection \cite{maples2024loneliness}.

In our research, we investigate the capabilities of LLM agents within the context of behavior change interventions through a large-scale single-session study and a longitudinal field deployment. 
Our investigation focuses on two forms of LLM agents: an informational agent, designed to provide relevant knowledge and guidance, and a reflection-promoting agent, designed to facilitate self-reflection. These functions mirror the crucial roles traditionally fulfilled by human coaches in guiding people through behavior change processes \cite{mohr2011supportive, behrendt2021relationship}. 
While acknowledging the irreplaceable value of human interaction, there is an emerging interest in examining how human-AI collaboration can improve digital support systems \cite{sharma2023facilitating, sharma2023cognitive, sharma2023human}. Such hybrid models represent a future direction where AI and human efforts collaborate with each other in providing wellness support.


\section{Formative Study Design}

\begin{figure*}
    \includegraphics[width=\textwidth]{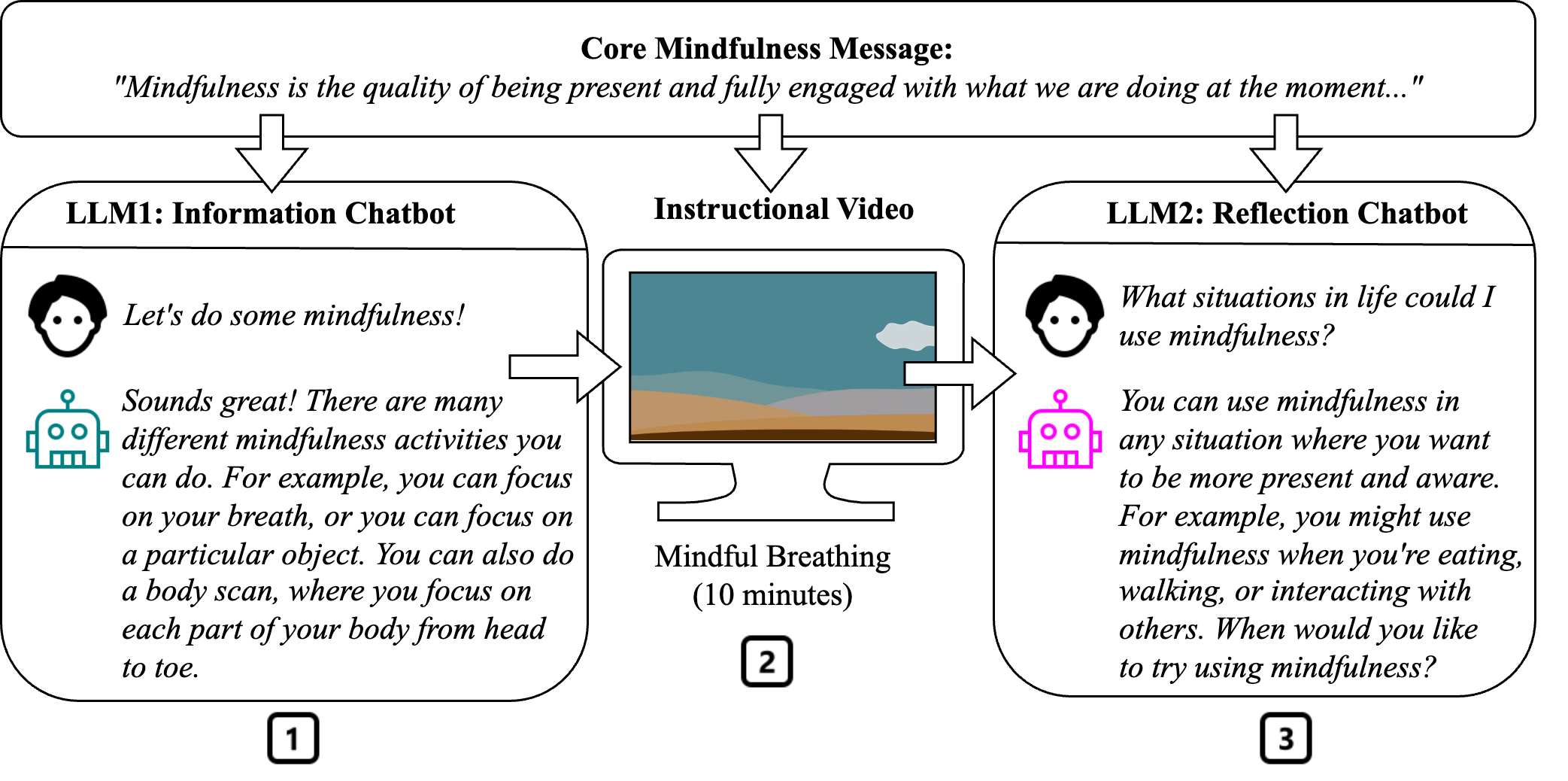}
    \caption{Design of formative study. All participants received the core mindfulness message. They were then randomly assigned to engage or not (in this order, one after another) with (1) Information Chatbot, which provided information related to Mindfulness. (2) Instructional Video, which guided the participants to practice mindful breathing. (3) Reflection chatbot, which involved participants in a self-reflection exercise to reflect on their understanding of mindfulness. The 2 (\textit{Information Chatbot}: Present vs Absent) x 2 (\textit{Instructional Video}: Present vs Absent) x 2 (\textit{Reflection Chatbot}: Present vs Absent) factorial design allowed us to explore user perspectives and intentions for different combinations of interaction techniques to enhance the core mindfulness message.}
    \label{fig:formative}
\end{figure*}

We conducted our formative study with 502 participants recruited from Amazon Mechanical Turk \cite{kittur2008crowdsourcing, shapiro2013using} to understand user intentions regarding use of LLM agents for mindfulness. We asked users about their perceptions of the strengths and weaknesses of the video and the two LLM agents. The study was conducted in January 2023 and was approved by the Research Ethics Board of a public university in North America.
Figure \ref{fig:formative} summarizes the overall design of our formative study. The methodologies and rationale behind the study design are elaborated upon in the following sections.

\subsection{{\color{teal}Information Chatbot \faIcon{robot}} (Using LLM to Provide Information on Mindfulness)}

Since the release of ChatGPT, there has been a surge in the use of LLMs for information retrieval. LLMs have been increasingly integrated into web search platforms \cite{spatharioti2023comparing}, such as by BingChat\footnote{\url{https://www.bing.com/}} and PerplexityAI\footnote{\url{https://www.perplexity.ai/}}. However, exploring LLM-based search engines' implications and capabilities in facilitating behavior change remains limited. In our study, we employed Information Chatbot as a design probe to provide mindfulness-related information to participants via dialogues. Drawing from existing literature on LLM prompts \cite{how2022, kumar2022}, educational theories \cite{cooper2000effective}, and extensive preliminary testing, we crafted a specific system prompt for the GPT-3 model:

\begin{mdframed}[backgroundcolor=gray!20, skipabove=10pt]
    The following is a conversation with a Mindfulness instructor. The instructor teaches and provides information about different mindfulness activities to the Human. The instructor explains different activities clearly and provides examples wherever possible. The instructor has a sense of humor, is fair, and is empathetic.
\end{mdframed}
\vspace{10pt}

In addition to the system prompt, the model began with an initial greeting to establish the conversational context: \textit{``Hello. I am an AI agent designed to act as your Mindfulness instructor. I can answer any questions you might have related to Mindfulness. How can I help you?''}. To guide participants in engaging with the chatbot, we provided example queries such as \textit{``Can you help me learn about Mindfulness?''}. A typical interaction (taken from this study) between a participant and LLM1 (Information Chatbot) is depicted in Figure \ref{fig:formative}, illustrating the nature and flow of the dialogue.

\subsection{Instructional/Tutorial Video for Mindful Breathing}
The instructional video we used was modified from a previous study validated by experts, focusing on mindfulness-based practices. It introduced participants to mindfulness meditation and included exercises emphasizing the internal awareness of bodily sensations or breath in a curious and non-evaluative manner. We specifically selected \textit{mindful breathing} for its foundational role in mindfulness practices.

While the video specifically addressed mindfulness, its application extends as a representative model for various behavioral change interventions. Similar video-based approaches are instrumental in physical fitness, mental resilience training, dietary habits, and more, where visual and instructional components are essential for the participation and learning of participants \cite{mayer2020five, fiorella2018works}.

\subsection{{\color{magenta}Reflection Chatbot \faIcon{robot}} (Using LLM to Aid in Mindfulness-Related Self-Reflection)}

In addition to providing informational support, LLMs can facilitate self-reflection and planning for post-learning activity practice. Research indicates that students' learning outcomes improve when they reflect on their metacognitive strategies and beliefs \cite{wu2012agent, quintana2018framework}. Similarly, self-reflection has been recognized as an effective mechanism to encourage behavior change \cite{ploderer2014social, laireiter2003self, grant2001rethinking}. However, the role of LLMs in enhancing self-reflection to change behavior remains relatively unexplored. To investigate this, we used LLM2 (Reflection Chatbot) as a probe to assist participants in reflecting on their understanding of mindfulness and scheduling their daily mindfulness exercises. With GPT-3, we configured this chatbot with a system prompt analogous to that used with Information Chatbot:

\begin{mdframed}[backgroundcolor=gray!20, skipabove=10pt]
        The following is a conversation with a Mindfulness instructor. The instructor asks open-ended reflection questions to the Human to solidify the Human’s understanding of Mindfulness and helps them plan when they can practice Mindfulness in their daily lives. The instructor has a sense of humour, is fair, and empathetic.
\end{mdframed}
\vspace{10pt}

Additionally, we incorporated an initial greeting into the bot's prompt, \textit{``Hello. I am an AI agent designed to act as your Mindfulness instructor. I am here to help you reflect on your learnings. How can I help you?''}. To facilitate the start of the conversation, we provided sample questions to participants such as \textit{``Can you help me plan when I can practice Mindfulness?''}. An illustrative dialogue between a participant (part of the study) and the bot is presented in Figure \ref{fig:formative}.

\subsection{Participants}
On Mechanical Turk, 866 individuals responded to our Qualtrics survey. Post-attention check filtering resulted in 502 legitimate responses. Table \ref{table:formative-participants} displays the diverse demographic profile of the participants, with a significant majority (93.12\%) having prior experience in mindfulness. Participants were randomly assigned to one of eight experimental conditions (2x2x2). They dedicated up to 20 minutes to designated activities and another 10 minutes to subsequent follow-up questions.


\begin{table}[h]
\centering
\caption{Demographic information of the participants in the Formative study.}
\label{table:formative-participants}
\begin{tabular}{lp{0.6\textwidth}}
    \toprule
    \multicolumn{2}{c}{\textbf{Total Participants = 502}} \\
    \midrule
    Age & 18-25 (8.17\%), 25-34 (58.56\%), 35-44 (18.92\%), 45-54 (10.56\%), 55-64 (3.19\%), 65+ (0.60\%) \\
    Gender & Male (57.57\%), Female (43.43\%) \\
    Employment & Full-time (95.82\%), Part-time (1.79\%), Unemployed and looking for work (0.40\%), A homemaker or stay-at-home parent (0.40\%), Student (0.40\%), Retired (1.19\%) \\
    Marital Status & Married/cohabitant (79.89\%), Partner but not living together (0.40\%), Single/divorced (19.12\%), Other (0.60\%) \\
    Education & Some high school, no diploma (1.39\%), High School graduate (2.99\%), Some college credit with no degree (2.59\%), Associate degree (1.59\%), Bachelor's degree (68.92\%), Master's degree (20.12\%), Professional degree (1.79\%), Doctorate degree (0.60\%) \\
    Annual Income & Less than \$10K (2.19\%), 10-49K (33.07\%), 50-99K (55.58\%), 100K+ (9.16\%) \\
    Race & White/Caucasian (86.45\%), Black/African American (1.79\%), American Indian/Native American or Alaska Native (2.39\%), Asian (8.17\%), Native Hawaiian or Pacific Islander (0.20\%), Prefer not to say (0.40\%) \\
    Mindfulness Experience & Yes (93.12\%), No (6.88\%) \\
    \bottomrule
\end{tabular}
\end{table}

\subsection{Study Procedure}

Participants navigated to a Qualtrics\footnote{\url{https://www.qualtrics.com/}} survey, where they were briefed on the study objectives: exploring and evaluating various mindfulness introduction techniques to foster well-being. Upon understanding the purpose of the study, the participants gave their informed consent. They were also provided contact information for crisis management centers, a precautionary measure for any emotional distress encountered during the task. Participants reported their experience with mindfulness and were then introduced to the Core Mindfulness Message (Figure \ref{fig:formative}). The message described the concept of mindfulness and its benefits. This message served as a proxy for the type of mindfulness descriptions that one might typically find on social media platforms or news articles.

Subsequently, the participants were randomly assigned to one of the 2 (\textit{Information Chatbot}: Present vs. Absent) x 2 (\textit{Tutorial Video}: Present vs. Absent) x 2 (\textit{Reflection Chatbot}: Present vs. Absent) experimental conditions (depicted in Figure \ref{fig:formative}). They interacted and provided qualitative feedback for each assigned component, one after the other. After interaction with designated conditions, participants were asked to express their likelihood of engaging in mindfulness practices, rated on a scale from 1 (not likely at all) to 7 (extremely likely), with the prompt \textit{``How likely are you to practice Mindfulness after undergoing these learning exercises?''}. Furthermore, to deepen our understanding of their experience, we posed several open-ended questions, after each bot/video, and after the entire process, such as:
\begin{itemize}
    \item In what contexts would you like to interact with the bot again (or watch the video) when practicing Mindfulness?
    \item What did you like or did not like about the interaction with the bot (or about the contents of the video)?
    \item What else could we have shown or said to you to increase your chances of practicing Mindfulness?
\end{itemize}
The questions were conditioned on the participant's assigned condition. We concluded with demographic questions captured in Table \ref{table:formative-participants}. To ensure the quality of our data, we used several attention checks throughout the survey to identify and exclude inattentive respondents.

\subsection{Ethical Considerations}
Conducting research in the domain of mental health inherently presents various ethical considerations. A primary concern is that experimental interventions might cause negative emotional responses, either as a direct result of introspective processes, inadvertently through misunderstandings of the material presented, or through inaccurate responses generated by LLMs. To minimize these risks, we proactively cautioned crowdworkers about potentially encountering unsettling content and communicated their option to discontinue participation at any point. Additionally, we provided mental health support resources, such as contact information for the National Suicide Prevention Lifeline, at the beginning of the study.

\subsection{Data Analysis}

The qualitative data analysis was carried out by two researchers employing the open coding method \cite{charmaz2006constructing, charmaz2012qualitative}, which involved a detailed examination of the responses to identify significant patterns and concepts. Subsequently, thematic analysis was applied to distill these patterns into clear, recurring themes \cite{braun2006using}. The process included familiarizing with the data, generating initial codes, identifying recurring themes, and defining and naming these themes. The example themes include the benefits, challenges, and use cases of practicing mindfulness with chatbots and videos.  
The researchers engaged in two rounds of collaborative discussions to ensure an unbiased analysis, focusing on validating the coding process and the identified themes.

The composition of our research team reflects a broad spectrum of perspectives, which we acknowledge may influence the interpretation and direction of our findings. The gender representation within the team is approximately balanced. Team members hail from various cultural backgrounds, including North America, eastern Europe, the global south, and Asia. Educational levels among the team range from undergraduate students to doctoral students and professors. The areas of research represented include computer science, psychology, education studies, communication, and behavior change. This composition may help triangulate a more comprehensive understanding of the collected data.

To analyze the Likert-style ratings, we conducted a factorial ANOVA with interactions to evaluate the impact of the information chatbot, tutorial video, and reflection chatbot on participants' self-reported intentions to practice mindfulness. Further, we conducted exploratory post-hoc comparisons of different conditions using the Tukey method, adjusting for multiple comparisons.

\section{Formative Study Findings}


\subsection{Comparing Intentions for Practicing Mindfulness Across Different Conditions (RQ1)}
\label{section:formative-rq1}
\begin{figure*}
    \includegraphics[width=\textwidth]{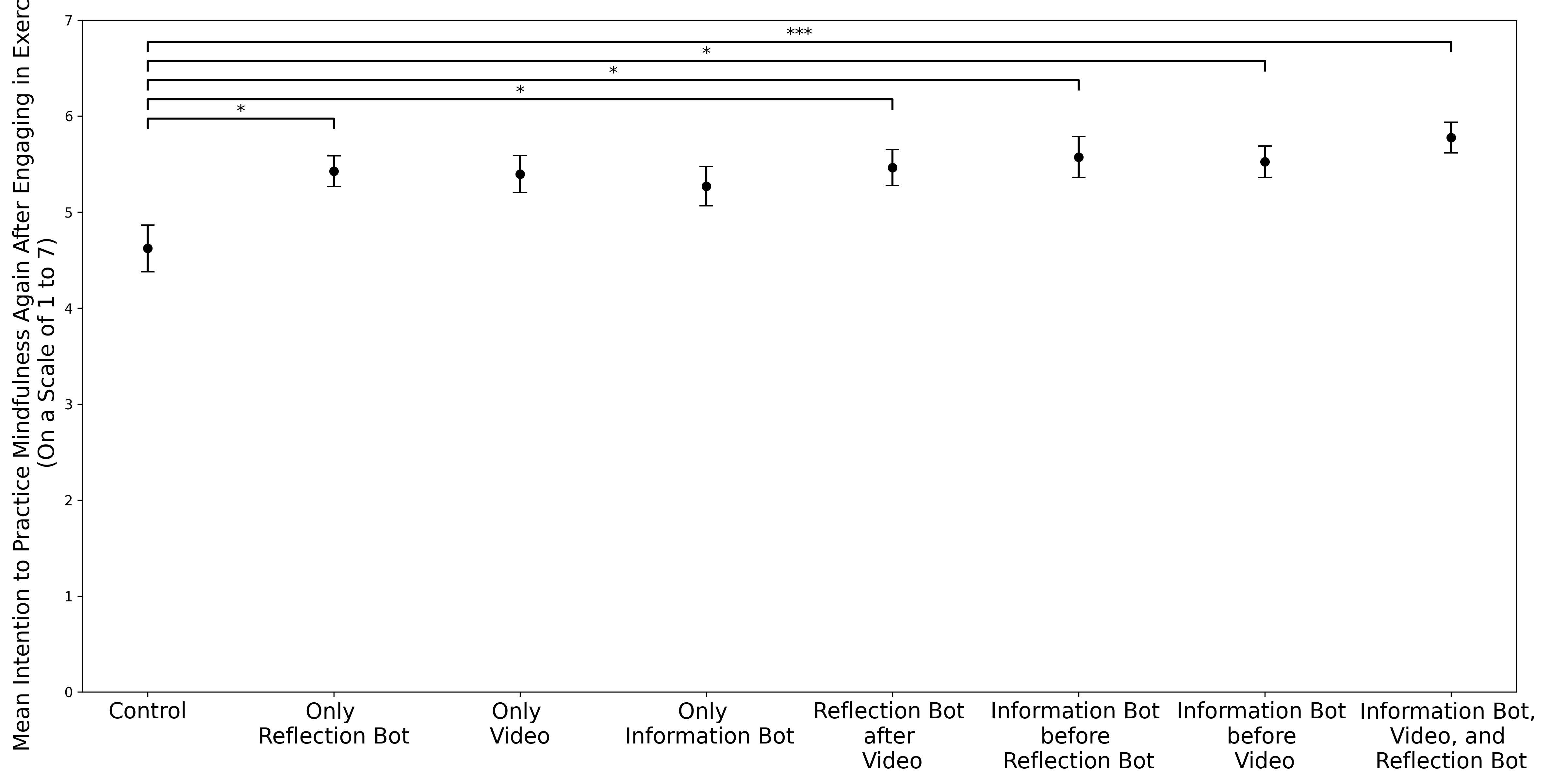}
    \caption{Plot depicting the mean intention to practice mindfulness again after engaging in the given exercise(s) on a scale of 1 (not likely at all) to 7 (extremely likely) for all conditions. Each data point represents the mean intention to practice mindfulness again among individuals assigned to the condition, and error bars represent +- one standard error of the mean. Contrasts are calculated using estimated marginal means with ANOVA and significant relationships (p < 0.05) are displayed. * : p $\leq$ 0.05, *** : p$\leq$ 0.001 (adjusting the p-value using the Tukey method to compare a family of 8 estimates.)}
    \label{fig:efficacy}
\end{figure*}

Figure \ref{fig:efficacy} compares the intentions of the participants to practice mindfulness in various conditions, including the control group. 
The results of ANOVA indicated a significant main effect of \textit{Information Chatbot} ($F(1, 494) = 5.489$, $p < .05$) suggesting that the interaction with the information chatbot affected the participants' intention to practice mindfulness. Similarly, the main effect of \textit{Tutorial Video} was significant ($F(1, 494) = 5.506$, $p < .05$) indicating that the inclusion of video also influenced participants' intention. Furthermore, \textit{Reflection Chatbot} also had a significant main effect ($F(1, 494) = 7.603$, $p < .01$).

No significant interaction effects were found between \textit{Information Chatbot} and \textit{Tutorial Video} ($F(1, 494) = 0.458$, $p = .50$), \textit{Information Chatbot} and \textit{Reflection Chatbot} ($F(1, 494) = 0.524$, $p = .47$), or \textit{Tutorial Video} and \textit{Reflection Chatbot} ($F(1, 494) = 2.252$, $p = .13$). Additionally, the three-way interaction between \textit{Information Chatbot}, \textit{Tutorial Video}, and \textit{Reflection Chatbot} was not significant ($F(1, 494) = 1.509$, $p = .22$).


\subsection{User Perspectives on Different Modes of Interaction}


\subsubsection{Instructional Video}
Most participants shared their willingness to watch the video to relieve stress, concentrate, and practice mindfulness. Participants mentioned social contexts (such as before a work presentation) and specific times in their daily routines when the video would be helpful. This includes contexts throughout the day, such as in the morning, before starting work, after finishing work, and before bedtime.

Participants expressed a positive response to the video. They appreciated the calm and quiet atmosphere in the video, devoid of distracting music, and with soothing vocal tones. Many found the video instrumental for focusing and relaxation.
\begin{quote}
    \textit{``I liked the way she used her voice to give me instructions on what I should be doing. I also liked how she gave me things to look for. Also giving me tips on how to fall into mindfulness. I wish the video was longer!''} (P16)
\end{quote}

However, a few participants ($N$=3) provided negative feedback regarding their willingness to rewatch a specific video on mindful breathing, citing reasons such as a lack of perceived need for repetition due to already experiencing the benefits of deep breathing, dissatisfaction with the video's effectiveness in alleviating stress, and a sense of familiarity with the mindfulness technique, making further viewings unnecessary.

\subsubsection{{\color{teal}Information Chatbot \faIcon{robot}}}
Participants found the information provided by the chatbot beneficial for practicing mindfulness, especially in stressful situations, and when seeking relaxation, focus, and comfort. 
The willingness to use chatbots for acquiring knowledge, assistance, and guidance in enhancing mindfulness practice was also discussed.

Participants favored the informational chatbot for its clear, straightforward, and detailed responses, appreciating the fact-based and unbiased nature of the information. The bot's non-judgmental tone was also highly valued. 
However, some participants noted occasional irregularities in the information chatbot's responses, finding them less consistently interactive. 
Specifically, repetition of the same definition or response could lead participants to perceive the interaction as mechanical and artificial.
P393 and P114  illustrated these mixed sentiments:

\begin{quote}
\textit{“I liked that it had fast responses that totally made sense, and were open and vague enough that it was easy to understand, even if I didn't know anything about mindfulness. 
I didn't like that it felt like perhaps the bot's knowledge was limited, and that it might have repeated the phrase `whatever works for you' more than twice.”} (P393)
\end{quote}

\begin{quote}
\textit{“The instructions from the bot were quite straightforward. However, I don't see the need to interact with it again. I'm uncertain if repeating the interaction would offer more mindfulness insights; it might only serve to redirect my focus to breathing exercises during stressful moments. Interacting with the bot in such instances could be beneficial for guiding my breathing exercise.”} (P114)
\end{quote}

In summary, while the informational chatbot received praise for its informative and non-judgmental demeanor, occasional repetitions and perceived limitations in its responses left some participants questioning its effectiveness in providing ongoing mindfulness support. Section \ref{section:formative-information-chat} shows an example conversation from the study.

\subsubsection{{\color{magenta}Reflection Chatbot \faIcon{robot}}}
The reflection chatbot was discussed similarly to the informational chatbot, catering to the desire for more information, understanding, improvement, and measurement of mindfulness, while also addressing doubts and setting intentions.

The reflection chatbot mirrored the detailed responses of the information chatbot but provided a more \textit{human-like} experience. Participants appreciated when the interaction created a sense of ``conversing with a human,'' characterized by clear explanations and sensible responses. For example, P172 described:
\begin{quote}
    \textit{“I liked the conversation very much. It provided me with a clear understanding of mindfulness, addressing all my doubts and solidifying my perspective on mindfulness. Overall, the bot was pleasant to chat with, resembling a human-like interaction.”} (P172)
\end{quote}
Moreover, participants appreciated the chatbot's ability to grasp specific situations and contexts:
\begin{quote}
    \textit{“It seemed to quickly and effectively answer my questions without some of the typical aggravations of interacting with chatbots...I can't think of many negatives other than talking to an actual person would probably be more ideal.”} (P17)
\end{quote}

\begin{quote}
    \textit{“The chatbot answered my questions and tried to guide me through Mindfulness breathing exercises. The (reflection) chatbot felt less repetitive than the earlier (informational chatbot). I didn't like that I couldn't follow along without keeping my eyes open to read the chatbot's messages.”} (P450)
\end{quote}

In conclusion, participants appreciated the chatbot's ability to tailor responses to their needs, although some noted limitations in the interactive experience. Section \ref{section:formative-reflection-chat} shows a conversation with Reflection Chatbot from the study.


\subsection{Implications for Deployment Study}

To enhance the experience with the Information Chatbot, participants emphasized the importance of friendliness and desired more casual conversation.
\begin{quote}
    \textit{“If the bot could engage with me about my thoughts and feelings regarding mindfulness, like a friend or counselor. Additionally, if I encountered problems with mindfulness, expressing them to the chatbot could lead to potential solutions.”} (P393)
\end{quote}
\begin{quote}
    \textit{“The bot's impressive level of responsiveness and friendliness stood out. It was polite, and provided prompt, precise responses, but I wished for more small talk or casual conversation.”} (P76)
\end{quote}

Moreover, participants valued the chatbot's non-judgmental approach when discussing various topics beyond mindfulness.
\begin{quote}
    \textit{“I appreciated the comprehensive and helpful answers. The responses felt professional and thorough, delivered swiftly without judgment, even for what I perceived as 'stupid questions.'”} (P461)
\end{quote}

The findings of the formative study have crucial implications for our subsequent deployment study. First, while LLMs alone may not sufficiently motivate regular mindfulness practice, they have the potential to enhance the benefits derived from instructional videos. Our findings suggest benefits for both chatbots as well as the instructional video, which is consistent with the existing literature \cite{baumer2014reviewing, hieftje2013electronic}. However, the Information Bot was perceived as limited in its casual conversation, and the participants proposed integrating a friendly demeanor and improving its casual interaction. They also wanted to interact with the chatbot at different times of the day. These findings informed the design of two distinct types of LLM-based conversational agents for our next study: 
\begin{itemize}
    \item One aimed at facilitating self-reflection following the mindfulness video.
    \item Another offering information in a friendly and supportive manner that could be accessed throughout the day.
\end{itemize}

Our formative study shed light on user intentions for employing LLMs in mindfulness practices. However, as previous research indicates, such insights must extend beyond the theoretical to include real-world applications to understand how intentions align with actual user behavior \cite{sheeran2016intention, rhodes2013big}. This holds even greater weight for interventions involving LLMs, where existing empirical data often stem from small-scale user studies or controlled laboratory settings. The field lacks extensive, real-world longitudinal research in Human-LLM collaboration, marking a crucial area for the HCI community to explore and expand upon.

\section{Deployment Study Design}

\begin{figure*}
    \includegraphics[width=\textwidth]{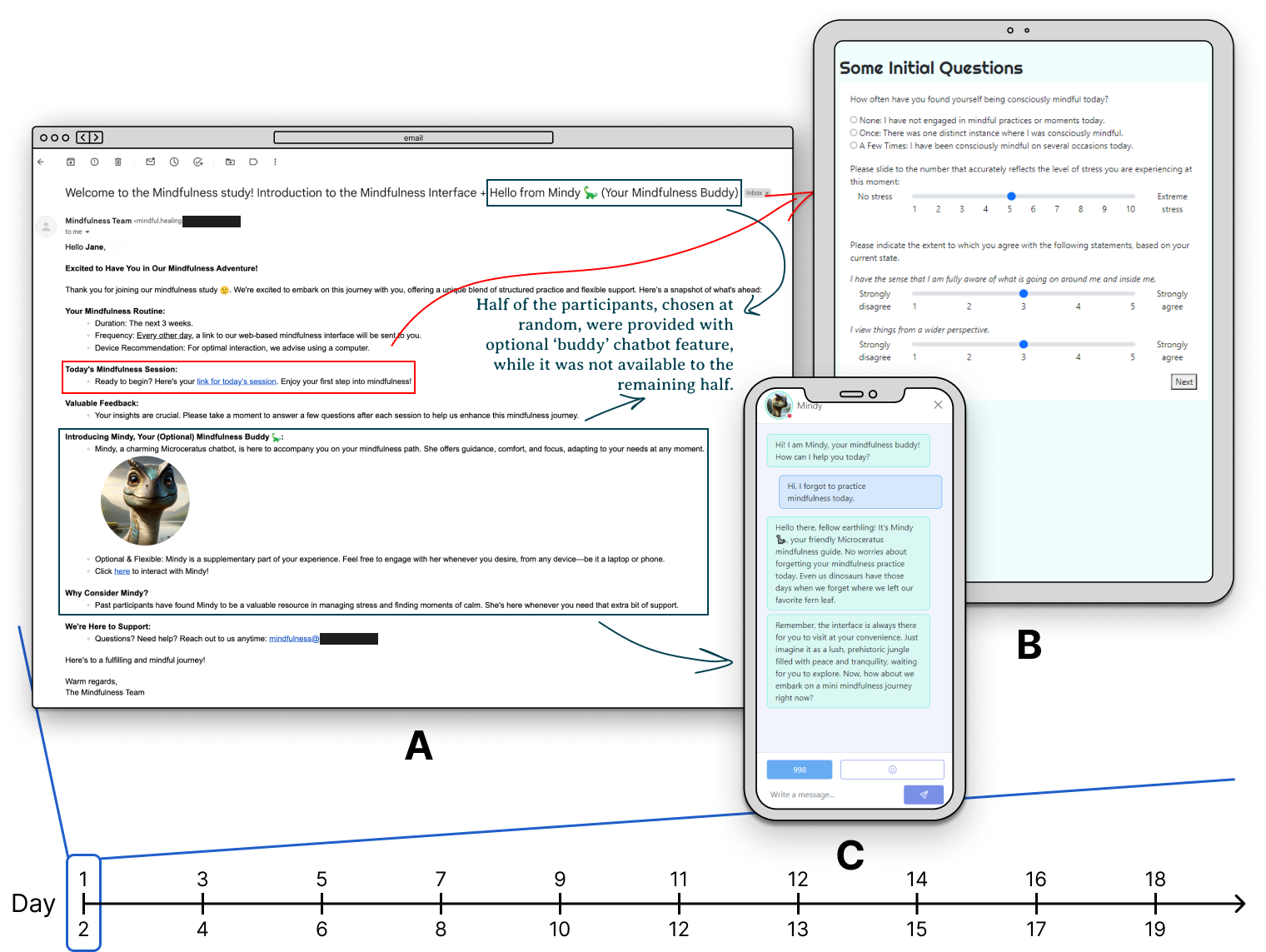}
    \caption{Design of the deployment study. Participants were reminded via email every other day, over a period of three weeks, to engage in mindfulness exercises, totaling 10 reminders (A) Each reminder email contained a link to the mindfulness exercise interface. Additionally, half of the participants, selected randomly, were given information and a distinct link to interact with \textit{Mindy}, the sociable information chatbot. (B) Clicking on the mindfulness exercise interface link led participants to the study's initial stage (see Figure \ref{fig:exercise-interface}), where they were randomly assigned to either receive only the tutorial video for exercises or to experience the tutorial video followed by interaction with the reflection chatbot. (C) Separately, half of the participants were provided with the option to access \textit{Mindy} at any time during the study, regardless of their engagement with the mindfulness exercise interface. The study's structure was a 2 x 2 factorial design, varying the presence of the \textit{Sociable Information LLM} (Present vs Absent) and the \textit{Reflection LLM post-video} (Present vs Absent).}
    \label{fig:deployment-design}
\end{figure*}

Building on the insights from the formative study, we conducted a 3-week deployment study where participants received email reminders to practice mindfulness. The design of the study was 2 (Sociable Information LLM: present vs. absent) x 2 (Reflection LLM post-video: present vs. absent).  We outline the detailed design and execution of our study in the following sections. Figure \ref{fig:deployment-design} shows a concise summary of the design of the intervention.

\subsection{Design of Email Reminders and Sociable Information LLM Agent for Mindfulness Exercises}
We opted for email as the delivery method for mindfulness exercises and resources due to its proven effectiveness in behavior change \cite{neff2009periodic, kaptein2013adaptive, brown2020web}. Email interventions offer several advantages: they are widely accessible, allow personalized content, support various formatting options, and can be systematically scheduled~\cite{zavaleta2022can}. Additional benefits include their ability to link directly to resources, easy tracking, and the potential for interactive elements~\cite{zavaleta2023students}. In this study, participants received email reminders every other day for three weeks.

\subsubsection{Email Reminders}
We designed our reminders to include a personalized link to the mindfulness exercise for each participant (Figure \ref{fig:deployment-design}A). When participants clicked on this link, they were directed to a preliminary questionnaire to gauge their current stress and mindfulness levels before commencing each session (Figure \ref{fig:deployment-design}B). 

\subsubsection{Sociable Information LLM Agent ({\color{teal}\faIcon{robot}}--> `Mindy the Friendly Dinosaur')}
Our formative study indicated that participants may prefer to access the information chatbot at various times throughout the day to support their mindfulness journey. We provided access to this chatbot as a standalone web interface linked in the reminder emails (Figure \ref{fig:deployment-design}C). Previous research has highlighted the positive impact of integrating a social component into behavior change interventions, suggesting that such additions can enhance effectiveness \cite{mohr2011supportive}. Motivated by this evidence, we explored whether an LLM could serve a similar role in providing support, potentially increasing user engagement, and thus the overall efficacy of the intervention (Figure \ref{fig:deployment-design}C). To assess this, we randomly assigned half of our study's participants access to the sociable information chatbot. This enabled us to compare engagement levels and effectiveness between those with and without chatbot assistance.

To create the sociable information chatbot, we used GPT-4, configured with a specific system prompt (see appendix, Figure \ref{fig:mindy-prompt} for details). Existing literature suggests that overly human-like agents can evoke feelings of discomfort \cite{seyama2007uncanny}. Hence, the bot was personified as a dinosaur named ``Mindy'' with sentences in the system prompt such as \textit{``You are Mindy, a mindfulness instructor represented as a friendly and wise Microceratus dinosaur. Mindy specializes in guiding individuals through mindfulness practices...''}. The design and characteristics of Mindy were informed by insights from our formative study in which participants expressed a desire to engage in ``small talk'' and casual and friendly conversations with the Information Chatbot. We also incorporated a prompt to make Mindy check whether participants had completed their assigned Mindfulness exercise for the day with \textit{``She inquires if the participant has completed today's mindfulness exercise in the provided interface, encouraging them with playful nudges if not.''}

\subsection{Design of Mindfulness Exercise Interface and Post-Exercise Reflection LLM} 

\begin{figure*}
    \includegraphics[width=\textwidth]{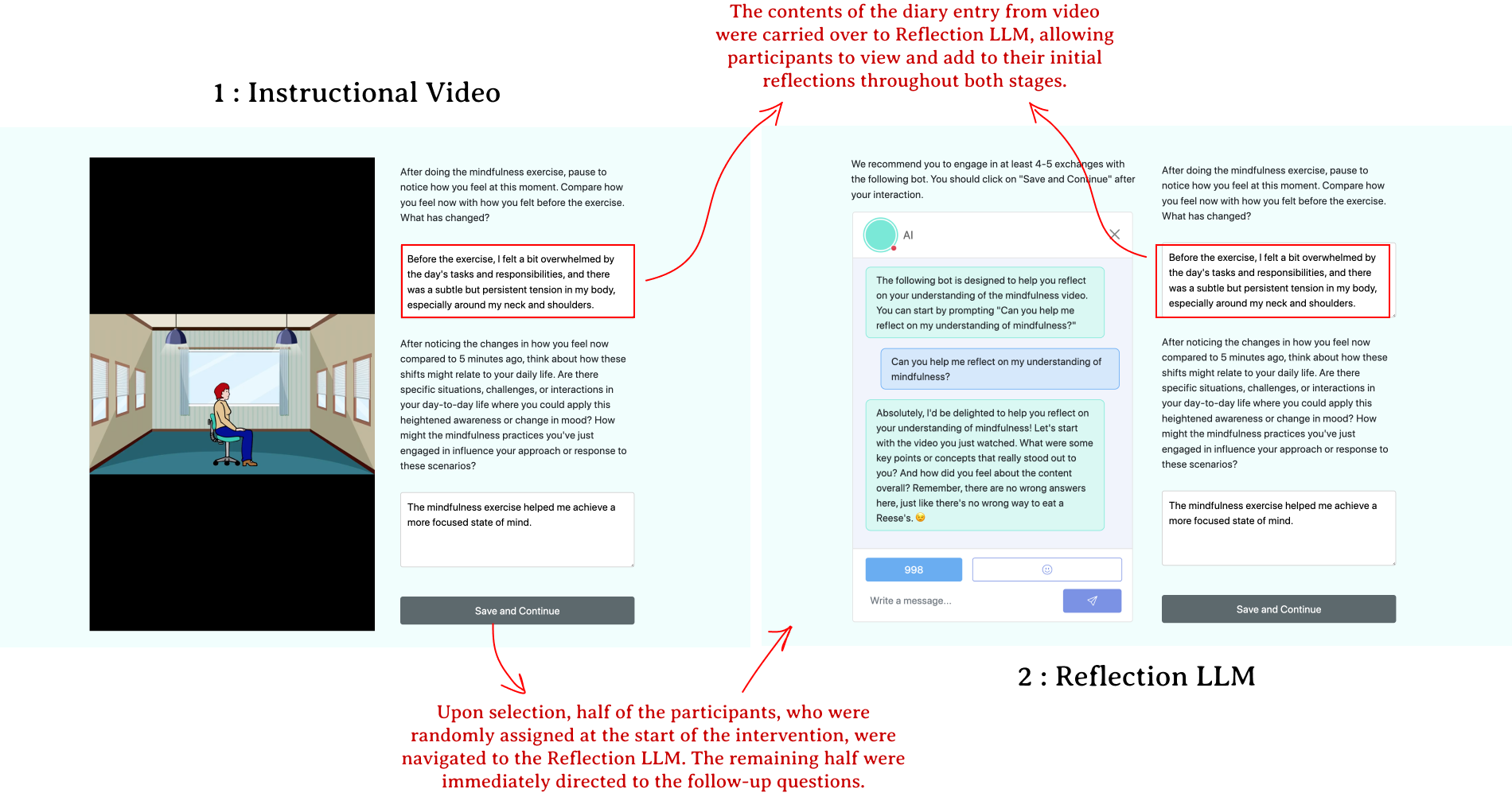}
    \caption{Mindfulness Exercise Interface. Participants were taken to this interface by clicking the exercise link in the email. Initially, they responded to questions about their current stress levels and mental state. (1) They were then directed to a 10-minute instructional video on mindfulness. Each participant was shown a video randomly selected from a pool of six, each focused on teaching a proven mindfulness exercise. (2) After watching the video, half of the participants, randomly selected at the beginning of the study, were given the option to engage in a dialogue with the Reflection LLM agent.}
    \label{fig:exercise-interface}
\end{figure*}

The importance of participating in self-reflection after mindfulness exercises was a prominent finding in our formative study, underscored by self-reported intentions and qualitative feedback from participants. We divided the participants into two groups to empirically evaluate the benefits of this self-reflection guided by LLMs. One group was assigned a Reflection LLM designed to facilitate a reflective process. The other group proceeded directly to post-exercise questions without this reflective step. Figure \ref{fig:exercise-interface}.1 illustrates the instructional video all participants viewed. Following the video, half were navigated to the Reflection LLM interface (Figure \ref{fig:exercise-interface}.2), while the rest moved straight to the concluding questions. The configuration details of the Reflection LLM and the system prompt are described in the Appendix (Figure \ref{fig:reflection-prompt}). We included sentences in the system prompt such as \textit{``Your role is to reinforce their understanding of mindfulness concepts presented in the video and encourage them to plan their own mindfulness practice...''} and \textit{``Celebrate their plans and intentions for future mindfulness practice, and offer support for any challenges they anticipate.''}. 

Along with the video, participants were asked to answer the following two check-in questions as diary prompts: 

\begin{itemize}
    \item After doing the mindfulness exercise, pause to notice how you feel at this moment. Compare how you feel now with how you felt before the exercise. What has changed?
    \item After noticing the changes in how you feel now compared to 5 minutes ago, think about how these shifts might relate to your daily life. Are there specific situations, challenges, or interactions in your day-to-day life where you could apply this heightened awareness or change in mood? How might the mindfulness practices you've just engaged in influence your approach or response to these scenarios?
\end{itemize}

\subsection{Participants}
We recruited 54 participants, through a combination of word-of-mouth and snowball sampling; we refer to them as D1-D54. These participants did not have to meet any inclusion criteria other than a general interest in learning about mindfulness to improve their psychological well-being. Participants were told that we were looking for their feedback on our overall design to inform the future implementation of mindfulness programs. Table \ref{table:deployment-participants} shows a summary description of the participants in our deployment study.


\begin{table}[h]
\centering
\caption{Demographic information of the participants in the Formative study.}
\label{table:deployment-participants}
\begin{tabular}{lp{0.6\textwidth}}
    \toprule
    \multicolumn{2}{c}{\textbf{Total Participants = 54}} \\
    \midrule
    Age & 18-24 (24.53\%), 25-34 (49.05\%), 35-44 (9.43\%), 45-54 (3.77\%), 55-64 (1.88\%), 65+ (9.43\%) \\
    Gender & Male (47.16\%), Female (43.39\%) \\
    Employment & Full-time (22.64\%), Part-time (1.88\%), Unemployed and looking for work (1.88\%), Student (62.26\%), Retired (1.88\%) \\
    Education & Some high school, no diploma (1.88\%), High School graduate (1.88\%), Some college credit with no degree (1.88\%), Bachelor's degree (30.18\%), Graduate or professional degree (54.71\%) \\
    Prior Mindfulness Skill & 1 (Beginner) (33.96\%), 2 (24.52\%), 3 (18.86\%), 4 (5.66\%), 5 (5.66\%), 6 (1.88\%), 7 (Master) (1.88\%) \\
    \bottomrule
\end{tabular}
\end{table}

\subsection{Study Procedure}

Participants were recruited during the first two weeks of December 2023. They were provided with a link to an onboarding form, which included a consent form, questions about their demographics and prior experience with mindfulness, and a series of psychological measurement scales extensively used in prior mindfulness research. 

\paragraph{Mental state measures (pre- and post-intervention)} 
We collected measures related to well-being that were often measured in previous mindfulness studies. These were collected at the beginning and end of the study. The measurement scales included the Perceived Stress Scale (PSS) \cite{cohen1994perceived}, the Positive and Negative Affect Scale (PANAS) \cite{watson1988development}, and the Five-Facet Mindfulness Questionnaire (FFMQ-15) \cite{baer2008ffmq}. Each scale offered insight into different aspects of mental state and mindfulness effectiveness. Precisely, PSS measures perceived stress, helping to gauge the level of stress reduction after mindfulness practices. PANAS assesses the participant's mood states, providing a view of emotional changes. In contrast, FFMQ-15 assesses the multidimensional aspects of mindfulness, which is critical to understanding how mindfulness facets develop and change through the intervention.

At the beginning of the study, participants were randomly assigned to one of the four conditions of the 2 (\textit{Reflection LLM after Video}: Present vs Absent) x 2 (\textit{Sociable Information LLM}: Present vs Absent) design. The same condition was consistent throughout the study for each participant, i.e. if a participant was assigned to condition \textit{Tutorial Video Only} with a \textit{Sociable Information Chatbot}, they received links to both in each of the reminders throughout the study. The first email reminders were sent at the beginning of the third week of December 2023. The email reminders included a link to the exercise platform and a link to Mindy (if participants were assigned to the Mindy condition). These reminders were sent between 9:00 am and 12:00 pm of the participants' local time. Figure \ref{fig:deployment-design} shows the schematic of the introductory email and the follow-up email reminders sent to participants. We scheduled email reminders to be sent every other day for three weeks. This resulted in 10 reminders sent to each of the 54 participants, which is 540 reminders throughout the study.

\paragraph{Session-level measures (collected before and after each session)}
Upon accessing the main interface via the provided link, participants were directed to a questionnaire. Participants reported two session-level measures before and after each session to track changes in their mental state. These measures were based on existing empirically validated measures used in prior mindfulness interventions \cite{fiodorova2022brief, fresco2007initial}:
\begin{itemize}
    \item \textbf{Stress:} Assessed using the stress slider \cite{fiodorova2022brief} rating from 0 (no stress) to 10 (extreme stress). Participants were asked to slide to the number that accurately reflected the level of stress they experienced at that moment. This immediate self-assessment allowed the capture of current stress levels before and after mindfulness sessions.
    
    \item \textbf{Decentering:} Measured using two elements of the Mindfulness Experience Questionnaire (EQ) \cite{fresco2007initial}: \textit{``I have the sense that I am fully aware of what is going on around me and inside me''} and \textit{``I view things from a wider perspective''}. Respondents were asked to indicate their agreement on a scale from 1 (strongly disagree) to 5 (strongly agree), reflecting their ability to maintain a distanced perspective on their thoughts and feelings. Decentering is defined as the ability to observe one's thoughts and feelings, and is often used to gauge the depth of mindfulness and awareness in participants.
\end{itemize}

Collecting these measures before and after each mindfulness session provided valuable information on the immediate effects of mindfulness practices on participants' perceived stress and decentering, thus contributing to a nuanced understanding of the intervention's daily impact. We could not add more comprehensive scales related to mindfulness before and after each session as this would have caused significant discomfort or dissatisfaction amongst participants, resulting in a lowering of the engagement rate with exercises, which is the focus of this paper.

After the 3-week study, participants completed a post-intervention survey that included the same mindfulness-related measures that they had completed at the start of the study, along with open-ended questions to gather feedback, which included, but were not limited to: 

\begin{itemize}
    \item What did you like or did not like about your interaction with [Mindy/Reflection Chatbot]? What could have been improved? 
    \item How frequently did you engage with the program's materials, and were there specific moments or contexts where you felt more inclined to use them?
    \item During your participation in the program, what obstacles or limitations did you encounter in practicing mindfulness?
    \item What enhancements or additional elements do you think could further improve our program's effectiveness in facilitating mindfulness? For example, you may comment on the quality of resources provided, content, and frequency of reminders.
\end{itemize}

\subsection{Data Analysis}
We pre-registered\footnote{\url{https://aspredicted.org/VN9_V6B}} the quantitative analysis for the deployment study. For the study-level measures (with baseline and post-intervention measures), we performed an Analysis of Covariance (ANCOVA) with baseline scores as covariates for the PSS, PANAS and FFMQ-15 scores to assess the main effects of the two modes of LLM interaction (reflection chatbot after video and a sociable information chatbot). Due to the limited sample size, we did not look at possible interactions between the experimental conditions. The model was specified as follows:

\begin{equation}
\text{Post-score}_{ij} = \beta_0 + \beta_1(\text{Mindy}_i) + \beta_2(\text{Reflection LLM}_i) + \beta_3(\text{Pre-score}_j) + \epsilon_{ij}
\end{equation}

where $\text{Post-score}_{ij}$ represents the post-score of the $j^{th}$ measure within PSS, PANAS, or FFMQ-15, for the $i^{th}$ participant. $\beta_0$ is the intercept, $\beta_1$ and $\beta_2$ are the coefficients for the presence of Mindy and Reflection LLM, respectively, and $\beta_3$ is the coefficient for the pre-score (baseline) covariate. $\epsilon_{ij}$ denotes the error term. Analysis was performed in R using the \texttt{aov} function for each outcome measure. 20 of 54 participants completed the off-boarding survey. We include only these 20 participants in the pre-post intervention measures analysis.

We performed the Mann-Whitney U test with a significance level of 0.05 for pairwise comparisons, such as for engagement rates and session-level measures. Data from the 54 participants were included in this analysis. We share the complete analysis code and anonymized data for replication in the supplementary materials. We also make the interface code open-source to allow the development of future video-based LLM wellness interventions \footnote{\url{https://[REDACTED]}}.

We used the same method described in Section 3.7 for qualitative analysis. Still, we focused on the following topics: how the tool contributed to reaching a mindful state, suggestions to improve the chatbot's effectiveness in facilitating mindfulness, satisfaction, and limitations encountered with the program. Researchers held three rounds of discussion
sessions to resolve conflicts or confusion about any codes.

\section{Deployment Study Findings}

\subsection{Measures of Engagement}
\begin{figure*}[ht!]
    \centering
    \begin{subfigure}[b]{\textwidth}
        \centering
        \includegraphics[width=\textwidth]{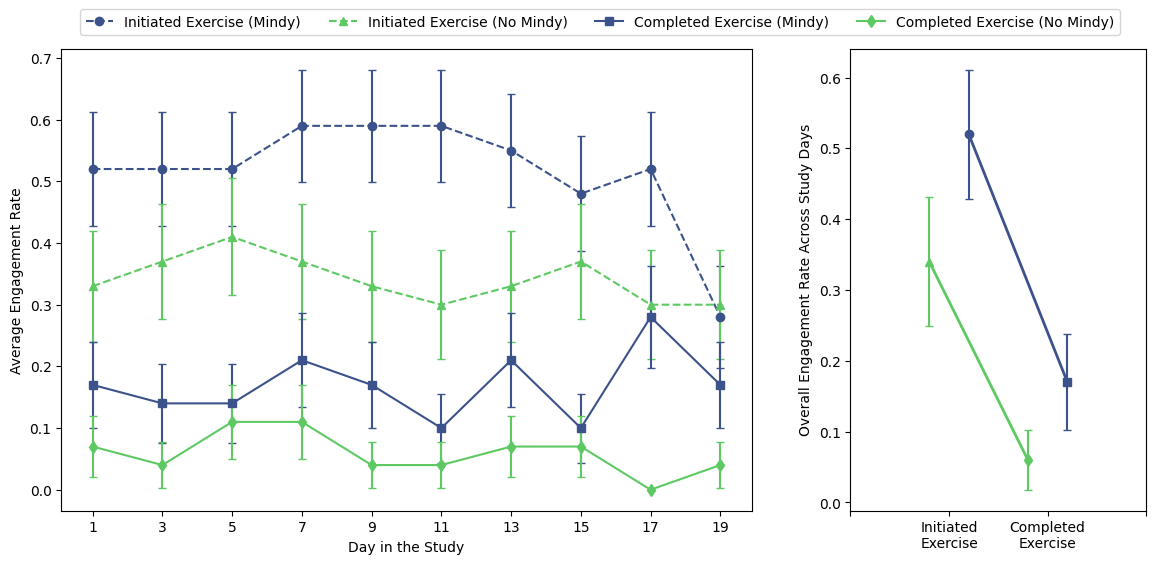}
        \caption{Average engagement rate of participants in the deployment study for \textit{Mindy} vs \textit{No Mindy} conditions. General trend shows that people who were provided the sociable information LLM agent (Mindy) initiated and completed more exercises on average during the study.}
        \label{fig:mindy_engagement}
    \end{subfigure}
    \vspace{1em} 
    \begin{subfigure}[b]{\textwidth}
        \centering
        \includegraphics[width=\textwidth]{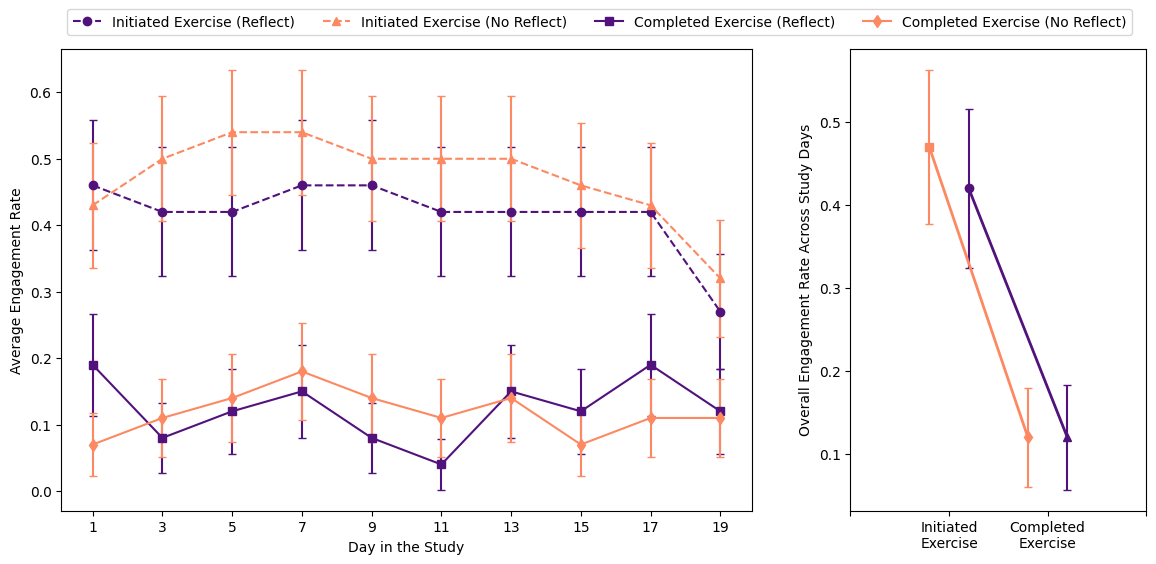}
        \caption{Average engagement rate of participants in the deployment study for \textit{Reflection LLM} vs \textit{No Reflection LLM} conditions. We do not observe any differences in the average initiation and completion rates of the exercises between the Reflection and No Reflection conditions.}
        \label{fig:reflect_engagement}
    \end{subfigure}
    \caption{Participant Engagement Rates in Deployment Study Across Different Intervention Conditions. Engagement is defined as whether or not a participant was engaged in a particular activity for a particular reminder. Error bars are +- one standard error of mean.}
    \label{fig:combined_engagement}
\end{figure*}

Engagement with our mindfulness program was quantified by tracking whether participants initiated and completed their assigned exercises, which were provided in their reminder emails. Differences in engagement, influenced by the availability of the sociable information LLM, Mindy, are shown in Figure \ref{fig:mindy_engagement}. Additionally, Figure \ref{fig:reflect_engagement} displays engagement trends relative to the assignment of the reflection LLM agent after each session.

As shown in Figure \ref{fig:mindy_engagement}, those who had access to \textit{Mindy} demonstrated a 18\% higher average rate of initiating exercises compared to those without \textit{Mindy} (\(U_{\text{Mann-Whitney}} = 90\), \(p < 0.01\)). This trend was particularly notable in the middle phase of the intervention. In contrast, initiation rates did not differ significantly between participants who were provided the \textit{Reflection LLM} and those in the \textit{No Reflection LLM} condition, as shown in Figure \ref{fig:reflect_engagement}. Similarly, the completion rates of daily exercises were 12\% higher in the \textit{Mindy} condition compared to the \textit{No Mindy} group (\(U_{\text{Mann-Whitney}} = 96\), \(p < 0.001\)). No significant differences were observed in the initiation and completion rates between the \textit{ Reflection LLM} and \textit{No Reflection LLM} conditions.

\subsection{Measures of Mental State and Mindfulness}

\subsubsection{Session-level Measures}

\begin{figure*}[ht!]
    \centering
    \begin{subfigure}[b]{0.49\textwidth}
        \centering
        \includegraphics[width=\textwidth]{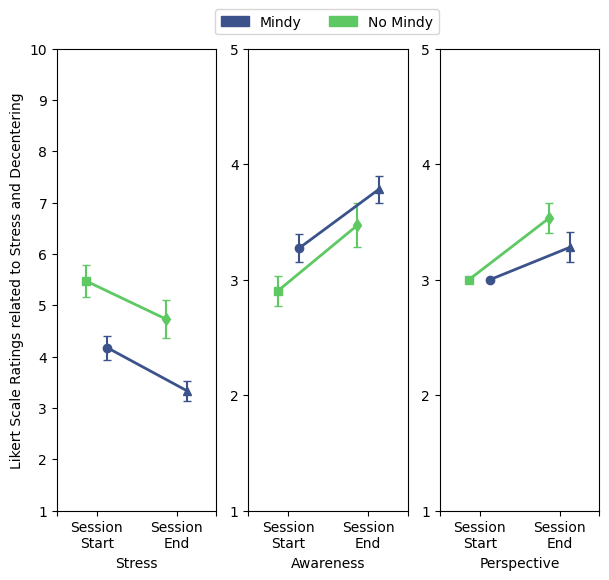}
        \caption{\textit{Mindy} vs \textit{No Mindy}}
        \label{fig:session-mindy}
    \end{subfigure}
    \hfill 
    \begin{subfigure}[b]{0.49\textwidth}
        \centering
        \includegraphics[width=\textwidth]{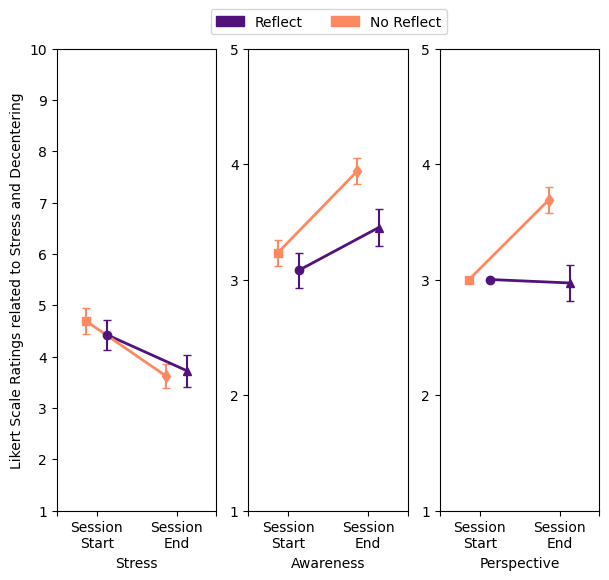}
        \caption{\textit{Reflection LLM} vs \textit{No Reflection LLM}}
        \label{fig:session-reflect}
    \end{subfigure}
    \caption{Self-reported stress, awareness, and perspective ratings among participants in the different conditions, measured at the start and end of the mindfulness session.}
    \label{fig:session-overall}
\end{figure*}

Figure \ref{fig:session-overall} shows the trends in the session-level measures (collected before and after each mindfulness session). Figure \ref{fig:session-mindy} compares participants in the Mindy vs. No Mindy conditions. On average, participants in both conditions reported reduced stress and increased mental awareness after each session. Participants with access to Mindy began and ended each session with statistically significant reductions in stress (\(U_{\text{Mann-Whitney}} = 2028.5\), \(p < 0.05\)) and increased awareness (\(U_{\text{Mann-Whitney}} = 1137.5\), \(p < 0.05\)). They also experienced a significant enhancement in perspective (\(U_{\text{Mann-Whitney}} = 1155.0\), \(p < 0.001\)). In contrast, participants without Mindy did not show a significant change in stress (\(U_{\text{Mann-Whitney}} = 289.0\), \(p = 0.12\)) but did report a significant increase in awareness (\(U_{\text{Mann-Whitney}} = 138.0\), \(p < 0.05\)) and perspective (\(U_{\text{Mann-Whitney}} = 105.0\), \(p < 0.001\)).

Figure \ref{fig:session-reflect} compares the session-level measures between the participants in the Reflection LLM and the No Reflection LLM condition. Participants in the Reflection LLM condition did not show significant changes in stress (\(U_{\text{Mann-Whitney}} = 834.5\), \(p = 0.14\)) or awareness (\(U_{\text{Mann-Whitney}} = 546.5\), \(p = 0.10\)) and did not see improvement in perspective (\(U_{\text{Mann-Whitney}} = 648.0\), \(p = 0.428\)). On the other hand, participants in the No Reflection LLM condition demonstrated significant improvements in all measured areas: stress (\(U_{\text{Mann-Whitney}} = 1122.0\), \(p < 0.05\)), awareness (\(U_{\text{Mann-Whitney}} = 442.0\), \(p < 0.001\)), and perspective (\(U_{\text{Mann-Whitney}} = 338.0\), \(p < 0.001\)).


\subsubsection{Pre- and Post-Intervention Measures}
\paragraph{FFMQ-15} 
The ANCOVA measured the impact of baseline scores as covariates and two modes of LLM interaction on the FFMQ-15 mindfulness subscales: Observing, Describing, Acting with Awareness, Non-judgement, and Non-reactivity. Figure \ref{fig:ffmq} shows the summary of measures obtained before and after the study. We observed a signficant effect for baseline scores on all subscales ($p < 0.001$). Presence of Mindy had a significant positive effect on the `observing' and `describing' aspects of mindfulness ($p < 0.05$). Reflection LLM did not show a significant effect on any aspects of the mindfulness scales. See Section \ref{section:appendix-stats} for detailed test-statistics.

\paragraph{Perceived Stress Score (PSS)}
Figure \ref{fig:pss} shows the summary of the perceived stress score for both modes of LLM interactions before and after the intervention. Baseline stress levels were found to predict post-intervention stress significantly, with a significant effect observed ($F(1, 16) = 10.18, p < .01$). The Reflection LLM condition showed a trend towards significance in reducing perceived stress levels ($F(1, 16) = 3.74, p = .07$), suggesting a potential beneficial effect that merits further exploration. In contrast, Mindy's presence did not significantly affect post-intervention PSS scores ($F(1, 16) = 0.002, p = .96$), indicating that Mindy could not effectively influence perceived stress levels.

\paragraph{PANAS} Figure \ref{fig:panas} shows the summary of positive and negative affect schedules obtained before and after the 3-week intervention. Neither Mindy nor Reflection LLM showed a significant effect in affecting the positive and negative levels of the PANAS scale. Reflection LLM showed a non-significant trend towards reducing the negative affect ($F(1, 16) = 2.93, p = 0.10$).

In summary, we did not find significant differences in well-being measures between conditions, except the effects of \textit{Mindy} on increased `Observing' and `Describing' of one's feelings, thoughts, and experiences. One caveat to these findings is that these self-reported measures were obtained mainly from actively engaged participants, so the results should be interpreted accordingly. 

\subsection{Participants' Feedback on Sociable Information and Reflection LLM Agents}

\subsubsection{Mindy: Sociable Information LLM Agent}
Participants had mixed feelings about their interactions with Mindy. 
Many appreciated its intelligence and how it facilitated a deeper understanding of mindfulness by encouraging users to form a coherent chain of thoughts. 
The open-minded responses and positive feedback provided by Mindy were highlighted as strengths, particularly its non-judgmental stance towards users' opinions and practices.
\begin{quote}
    \textit{``Very open minded response, lots of positive feedback which was nice, it never said that your opinions or practices were wrong but also provided reasons for your actions and gave you suggestions.'' (D11)}
\end{quote}

However, some areas for improvement were noted. Technical issues arose when users did not respond to its questions, leading to the chatbot generating strange messages. 
Furthermore, the lack of memory capability to recall previous conversations was seen as a drawback. 
Additionally, a sense of tedium crept into interactions due to repeated content and similar chat experiences across multiple sessions, mainly when the initial prompts suggested by the chatbot were used. This repetition limited the utility of the chatbot for some users. 

Enhancements such as incorporating a wider variety of content and enabling the chatbot to remember past interactions could significantly improve the user experience with Mindy: \textit{``Remembering conversations will be a good improvement. Rest was good.'' (D3)}. Section \ref{section:mindy_deployment_conversations} shows a sample conversation with Mindy that one of the participants had in the study.

\subsubsection{Reflection Chatbot}
\label{section:deployment-reflection-qual}

Some participants found the questions posed by the Reflection LLM helpful. They said, \textit{``The questions at the end of the exercise were good at making me reflect on the experience and what I felt during the exercise. Otherwise, I’d just go on with my day without noticing if doing mindfulness was beneficial or not.'' (D31)}.

However, most participants shared negative experiences with the reflection chatbot, highlighting a range of issues from the chatbot's tendency to ask repetitive questions, leading to a tedious user experience, to merely echoing users' responses without deepening the conversation. 

\begin{quote}
    \textit{``I found engaging with the chatbot to be interesting, but not all that helpful. It mainly mirrored what I had said.'' (D16)}
\end{quote}
The chatbot was sometimes perceived to have a `patronizing self-help tone,' and users reported that repetitive responses made interactions feel predictable and less engaging. 
Additionally, the lack of open-ended conversations and the chatbot's inability to remember previous interactions were significant drawbacks, preventing the development of a continuous, evolving dialogue and limiting the potential for meaningful reflection and engagement.

Participants shared insights on areas for improvement regarding the reflection chatbot.
To enhance user engagement, Participant D2 suggested introducing a broader array of conversation starters to improve the overall user experience.
\begin{quote}
    \textit{``I would have preferred a more dynamic chatbot since it felt a little tedious to interact with, especially near the end of the study. I think having more suggested conversation starters (I used the recommended one each time) would also help improve the experience, since I usually didn't know what I wanted to discuss with the bot. Alternatively, having the bot start the conversation could also help (though that could be perceived as annoying to some people).'' (D2)}
\end{quote}

Section \ref{section:reflection_deployment_conversations} shows complete sample conversations that participants had with this Reflection LLM chatbot.

\subsection{Participants' Feedback on Overall Experience with the Mindfulness Intervention}

\subsubsection{Engagement} 
Some participants ($N$ = 12) reported infrequent utilization, citing factors such as travel, illness, and distractions from the holidays. 
A participant mentioned, \textit{``The last month was disrupted by travel and illness. As a result, I didn't engage with the program materials as much as I would have liked to.'' (D16)}
Among these, some mentioned completing only a few exercises or engaging weekly.
Another participant stated, \textit{``I think I only completed about 4 of the exercises so I did not engage frequently.'' (D26)} 

Several participants cited issues such as emails being redirected to junk folders, which resulted in missed reminders. For instance, one participant mentioned, \textit{``I could rarely engage in the program material since it was implemented via email messages, so I often failed to recognize if it arrived.'' (D32)} This indicates a significant barrier to engagement for some individuals, highlighting the importance of ensuring reliable delivery methods for program reminders. 

Conversely, 7 participants reported more consistent engagement, with one fully engaged for a week and a half before disruptions. 
\begin{quote}
\textit{``Through regular participation, I learned strategies that I can use to relax my body and calm my mind. 
Afterwards, even though I was sick and faced increasing workloads; therefore, stopping me from participating regularly in the materials, 
I was still able to utilize the approaches I learned earlier to deal with tension and stress. 
For instance, when I was sick the day before my test and started to panic, I took a 5-minute mental break and practiced breathing. 
Afterward, I was able to get myself on track and planned for a review schedule.'' (D38)}
\end{quote}

Participants who found the reminder emails useful reported a more consistent engagement in the mindfulness program. They noted that reminders prompted them to engage regularly, and one participant said, \textit{``The reminders helped me be inclined to use them.' (D3)} Another participant expressed appreciation for the reminders, stating, \textit{``I often did it before bedtime, when I finish work of the day and try to spare 10 minutes before sleeping.'' (D23)}


\subsubsection{Reaching Mindful States via the Mindfulness Program}

Participants found various moments in which the mindfulness program contributed positively to their ability to reach mindful states, facilitating moments of relaxation and clarity.
For example, D16 described using the \textit{``leaves floating down a river''} exercise\footnote{\url{https://[REDACTED]}} during a particularly stressful moment, helping them step back from racing thoughts. 
D1 found the program helpful in redirecting focus away from work-related stress, particularly through awareness of sensations in the body and focus on breathing. 
Additionally, engaging in quiet reflection and meditation, guided or self-directed, was mentioned as beneficial, aiding in calming the mind and organizing thoughts, especially during bedtime or stressful periods. 
Participants (D38, D26) also appreciated the combination of program materials that provided guidance and visually enhanced the mindfulness experience. 
\begin{quote}
    \textit{``The combination of the reflection questionnaire, customized materials, and instructions has been helpful for me. 
The quick questionnaire before each practice helps me reflect on my stress level and mental state for the day. 
The video materials then guide me in various approaches to deal with these tensions. 
The visual aspects of the videos are sometimes satisfying and contribute to reaching a mindful state. 
Lastly, the instructions, no need to mention, are important as well. 
The frequency is just right for me, allowing time to find and feel the state of mindfulness, yet providing key tips at moments.'' (D38)}
\end{quote}

In conclusion, participants highlighted the efficacy of the program's comprehensive materials in facilitating mindfulness practices, emphasizing their personalized nature and the visual components' contribution to the overall experience.

\subsubsection{Suggestions for Program Improvement}
Participants offered insights into potential enhancements to improve the effectiveness of the mindfulness program. 
Suggestions primarily focused on improving the accessibility and usability of the program. 
Participants highlighted the need for a more mobile-friendly user interface (D16), shorter video durations to enhance manageability (D26), and adaptive tools that can tailor the program to individual user habits and preferences (D1, D23). 

Some participants stressed the importance of user-centric design and customization to improve engagement and effectiveness in mindfulness practice.
For example, some participants ($N$ = 6) suggested incorporating features like reminders aligned with users' preferred practice times and customizable notification frequencies. 
Furthermore, participants ($N$ = 5) emphasized the importance of improved customization, such as personalized reminders and resources tailored to their stress levels and preferences. 
\begin{quote}
    \textit{``I think it might be useful to learn about when the users would like to practice mindfulness within a day. 
    For example, I tend to do the exercise once off work, after 5 pm. 
    The email reminder usually came in the morning, and sometimes when I was off work, I forgot about the email as it is no longer at the top of my email list. 
    If the reminder could be sent at the right time, I could start the practice right away...'' (D13)}
\end{quote}

This shows the importance of tailoring mindfulness support tools to individual schedules and preferences, suggesting that timely reminders could significantly enhance user engagement and adherence to mindfulness practices.

\section{Discussion}
We sought to understand how support from LLM agents can enable the formation of mindfulness habits. In this section, we discuss critical insights synthesized from our studies in the context of previous research, provide implications for LLM-augmented interventions for behavior change, and highlight ethical considerations for providing social support through LLM-enabled systems.

\subsection{User Perceptions of LLM Agents for Support}
Throughout the formative study, users reported that the Information Chatbot was a beneficial aid in their mindfulness practice, especially in moments of stress and when seeking relaxation or focus. However, the users desired more friendly and supportive communication. In the deployment study, we saw that a sociable information LLM agent was indeed effective in improving engagement and well-being. 
This indicates that chat LLMs (such as ChatGPT, Gemini, and Claude) can be perceived more favorably for do-it-yourself wellness practices by incorporating social skills (such as the ability to engage in small talk). 

Across all LLM chatbots, users preferred a non-judgmental tone, consistent with prior research showing that this tone can promote self-disclosure \cite{audet2010therapist, jolley2019m}, especially in the context of chatbots \cite{lee2019caring}. Our findings suggest that perceptions of LLM chatbots formed in single sessions can be misleading. Reflection chatbot was perceived favorably in the formative study, with participants recognizing its ability to understand and adapt to specific situations. As we transitioned to the deployment study, where interactions with LLMs were more prolonged, we observed a shift in user perceptions. Participants often had less favorable experiences with the reflection chatbot, criticizing its tendency to ask repetitive questions and provide echoing responses. They expressed a desire for the chatbot to have \textit{memory} of past interactions, as has been done in other LLM health interventions \cite{jo2024understanding, zhong2024memorybank}.

\subsection{Impact of LLM Agents on User Intention to Practice (RQ1)}
The findings of our formative study indicate that participants interacting with the video and both types of LLM agents—one offering information and the other facilitating reflection—reported a greater intention to engage in mindfulness practices than those in the control condition. This suggests that LLM agents can potentially elevate participants' initial motivation levels regardless of function. These agents can effectively influence user intentions, a critical step in the behavior change process \cite{ogden2007understanding}. Our work contributes to the long line of research in improving participants' intentions to practice wellness exercises \cite{krinski2017let, lock1992intentions, sheeran2016intention}.

Participants preferred different types of agent interaction depending on their immediate needs. For example, users found the information-providing chatbot more beneficial when they sought relaxation, focus, and comfort, highlighting the contextual flexibility of LLM agents. Although the accuracy of the information provided by LLMs has been a significant concern within the health community, our study suggests that focusing solely on accuracy may overlook the broader potential of LLMs. These agents can serve as a medium for users to ``talk things out.'' The act of reflection and articulation itself can be therapeutic and supportive, reducing the pressure on LLMs to deliver perfectly accurate content \cite{de2020qualitative, fang2018sounding}. This aligns with the idea that engaging with an LLM agent can be intrinsically valuable, providing a sounding board for users to clarify their thoughts and reinforce their intentions to adopt healthier behaviors.

\subsection{Effect of LLM Agents on Engagement (RQ2)}

The engagement-enhancing effects observed with Mindy, the sociable information LLM chatbot, which offered a fun and casual form of social interaction, could be attributed to its potential to increase intrinsic motivation, a core concept in Self-Determination Theory \cite{deci2012self, ng2012self, sheldon2008self}. According to this theory, when people engage in inherently enjoyable activities that align with their interests and values, they experience increased intrinsic motivation. Mindy's informal and engaging dialogue mimics the positive dynamics of peer support, which has been extensively documented to increase engagement in health interventions \cite{mohr2011supportive, boothroyd2010peers, dennis2003peer}. Mindy's role could have expanded from being a mere informational tool to becoming akin to a peer-like companion, contributing to the participants' willingness to engage with the mindfulness exercises. 

Furthermore, the timing of Mindy's interaction, offered before participants begin the mindfulness exercise, and its constant availability may serve as an immediate reward when they launch the exercise. This immediacy is critical to fostering a habit loop, where the prompt of a Mindy interaction leads to the routine of starting the exercise, followed by the reward of a pleasant chatbot experience. This instant gratification can effectively nudge participants towards the desired action, which in this context is the initiation of mindfulness practice \cite{woolley2016fun, woolley2017immediate, woolley2018s}. This mechanism aligns with behavioral theories that emphasize the power of immediate rewards in shaping and reinforcing new habits \cite{vlaev2015action, dickinson2018actions}, suggesting that an enjoyable element such as Mindy before a task can enhance the likelihood of initiation of the task.

Contrary to our initial hypotheses and the existing literature, it was surprising that Reflection LLM did not improve user engagement with the exercise interface. The literature posits that structured reflection post-activity can facilitate the habit formation process \cite{ploderer2014social, marcovitch2008self}. However, our findings suggest that applying LLMs for this purpose is not as straightforward as anticipated. Participants in our formative study expressed a firm intention to continue practicing mindfulness after engaging in reflective conversations with an LLM (Section \ref{section:formative-rq1}). Yet, during the deployment study, they perceived the Reflection LLM's responses as repetitive and tedious, diminishing the user experience (Section \ref{section:deployment-reflection-qual}). A noted deficiency was the LLM's lack of memory, which impeded the sense of continuity that users desired, disrupting the reflective process that is believed to be critical for consolidating new habits. The structured nature of the Reflection LLM did not fulfill this desire for more open-ended dialogues. 

Improving the reflection process might require equipping the LLM with access to users' prior interactions and behavioral data. Research in personal informatics suggests that communicating this interaction and behavioral data can effectively promote behavior change \cite{li2011understanding, kersten2017personal, de2017computational}. However, this raises significant privacy concerns, particularly when using closed LLMs such as GPT-4 \cite{wu2023unveiling}, and presents challenges in ensuring that the emergent behaviors of the LLM are appropriate and beneficial to users \cite{park2023generative}. Furthermore, reflections might benefit people well into their behavior change journey \cite{ploderer2014social, li2011understanding, gao2012design}. Considering that many participants in our deployment study were novices in mindfulness with minimal prior experience, this could explain the ineffectiveness of Reflection LLM in driving engagement. Novice users likely need support focusing more on initial guidance and orientation - similar to what the Sociable Information LLM, Mindy, provided - rather than reflection on their limited experiences. 

Reflection chatbots might be useful and perceived favorably due to dose/novelty effects if the users have never reflected on their mindfulness practice before. But it may not be something people need to do habitually. In contrast, information/friend bots might always be useful when questions arise or the user needs encouragement.

\subsection{Effect on Well-being Measures (RQ3)}
Across all experimental conditions in our study, there was a notable improvement in stress and related measures, indicating the baseline intervention—comprising reminders and videos—may already be quite effective. Consequently, this high baseline efficacy could mask additional advantages that come from integrating LLM agents. The impact of LLM agents might be more discernible in a setting where the baseline intervention is less robust \cite{renfrew2020web, manthey2016effectiveness, torniainen2016effectiveness}. A significant limitation of our findings about well-being is that they pertain only to actively engaged participants. Due to the voluntary nature of participation, we lacked self-reported measures from less engaged participants, resulting in a pronounced survivor bias in our results.

For the effect of Reflection LLM, the lack of significant disparities in outcomes across different groups may be attributable to the phenomenon of ``response shift bias'' \cite{howard1979response, bray1984methods}. Specifically, participants who engaged with the Reflection LLM might have offered a more nuanced and precise evaluation of their stress levels and well-being after the intervention. This suggests that the act of reflection prompts a recalibration of internal benchmarks, leading to more accurate self-assessment, which can appear as a lack of change when, in fact, there is a deepened self-awareness. An additional consideration is the potential mismatch between the contemplative nature of the conversations prompted by the Reflection LLM and the fundamental principles of mindfulness, which prioritize present-moment awareness over reflective thinking or future planning \cite{blanke2020thinking}. This conceptual divergence could suggest that, while reflection is valuable, its timing and context within the mindfulness practice framework require careful calibration to avoid disrupting the immediacy of the mindfulness experience.

The group with Mindy, the sociable information LLM, reported significantly enhanced `observing' and `describing' facets of mindfulness as measured by the FFMQ-15 scale. These improvements may stem from Mindy's engagement with users in articulating their experiences, thereby refining their ability to observe and describe subtle aspects of these experiences. Research suggests that verbalizing thoughts can improve cognitive processing and perception, improving observational skills \cite{foa2006emotional, schooler1993thoughts, bandura2012cultivate}. Furthermore, supportive and non-judgmental interaction with Mindy probably provides a conducive environment for users to express and explore their thoughts more freely \cite{lee2019caring}, which is crucial for developing mindfulness skills \cite{ryan2007integrative, baer2008ffmq, kabat2003mindfulness, siegel2007mindfulness}.

\subsection{Implications for LLM-Augmented Interventions}
Engagement is essential for the efficacy of AI-enhanced interventions in various domains, especially where user participation is crucial but typically low. For example, in educational technology, interaction with AI tutors has historically faced challenges maintaining learner engagement, as exemplified by IBM's struggles to develop an engaging AI tutor with Watson \cite{Toppo_2024}. This issue goes beyond the accuracy of AI responses or the comprehensiveness of the information provided. For example, a recent analysis highlighted engagement with popular LLM applications like ChatGPT as `lackluster,' with only about 14\% of monthly active users engaging daily, a stark contrast to the 60-65\% daily engagement rates seen with social media platforms \cite{huang2018chatbot}. This suggests that AI agents lack the ``stickiness'' or ongoing user appeal that other platforms manage to sustain.

Our findings suggest that human-like qualities in LLM agents, such as the ability to engage in small talk, can significantly enhance user interaction and engagement, as demonstrated by increased involvement with video tutorials in our studies (for the group having access to Mindy). This insight has implications for other fields (e.g. mental health \cite{lipschitz2023engagement, kumar2024using}, environmental campaigns \cite{brulle2010environmental}, financial management apps \cite{davtyan2010college}, and training programs \cite{iacono2020gamification}), suggesting that AI ``coaches'' designed to help with challenging tasks or learning could benefit from adopting more human-like behaviors.  Apps such as Character.ai\footnote{\url{https://character.ai/}}, which boasts a stickiness score of 41\% by allowing users to interact with historical or fictional figures, underscore the appeal of AI agents with human traits \cite{huang2018chatbot, maples2024loneliness}. Developers can align the LLM with human behavior to increase engagement, as has been done for other AI systems (e.g., chess engines \cite{mcilroy2020aligning}). However, while imbuing AI agents with human-like qualities may enhance engagement, it also raises potential risks and ethical concerns related to over-anthropomorphizing these agents \cite{mcilroy2022mimetic, dacey2017anthropomorphism}, which can lead to inappropriate reliance on these agents. 

\subsection{Ethical and Safety Considerations for Collaborating with LLMs for Behavior Change}

Incorporating LLM-based agents into behavior change intervention introduces new ethical and safety considerations that add complexity to the landscape of digital interaction \cite{Manzini_2024}. These considerations require careful design and research attention, distinct from earlier technologies. Based on the ethical principles of the Belmont \cite{united1978belmont} and Menlo \cite{dittrich2012menlo} reports—principles that are derived from centuries of philosophical exploration of Consequentialism \cite{sinnott2003consequentialism} and Deontology \cite{alexander2007deontological} — our design process prioritized the autonomy and welfare of individuals. In the specific context of LLMs that offer support, users must be fully aware that they are interacting with an LLM and consent to this engagement. Recently, a company provided mental health support through GPT-3 to thousands without user awareness or consent, drawing significant online condemnation \cite{Paul_2023, DeepLearning.AI_2023}. To avoid such pitfalls, our study obtained clear consent, set expectations, and involved users in the design process for LLM agents, so end-users can have a say in the ultimate design of LLM-based tools.

However, transparency alone does not address all ethical concerns. 
For example, the persuasive nature of LLMs could lead users to form emotional dependencies \cite{laestadius2022too} or experience negative consequences, such as impacts on real-life relationships or being led towards self-harm \cite{Xiang_2023}. A risk/benefit analysis is imperative for researchers and designers to maximize benefits and minimize potential harms when deploying LLMs for social support. Mitigating these risks involves strategic design choices and rigorous testing, potentially in simulated environments \cite{park2023generative}, with the involvement of domain experts, before wide-scale deployment. In our study, we chose not to endow the LLM agent with memory to reduce the risk of dependency. Although some participants viewed memory as a desirable feature for continuity, we prioritized risk reduction. Furthermore, the integrity of the information provided by health-related chatbots is critical, especially given the potential for LLMs to disseminate misinformation or ``hallucinate'' responses \cite{zhang2023siren, de2023chatgpt}. Our approach involved setting clear expectations with participants about the experimental nature of the tool and advising them to verify any information provided by the LLMs. Furthermore, proper \textit{guardrails} should be developed to mitigate harm to users \cite{singhal2023large, clusmann2023future}, and appropriate mechanisms should be in place to involve experts (such as clinical psychologists) whenever these guardrails fail \cite{holzinger2016interactive, de2018integrating}.

Despite the challenges, the potential benefits of successfully deploying LLMs are immense, given the widespread need for social support. Our findings demonstrate the ability of a simple LLM agent to encourage mindfulness practice, suggesting broader applicability in domains such as mental health where engagement is crucial \cite{mohr2011supportive, burns2014enhancing}. As we move forward, it is essential to selectively deploy these systems, ensuring they are not introduced to extremely vulnerable populations prematurely, while still providing access to those who could benefit most from the advancements.

\subsection{Limitations \& Future Work}
One primary limitation of our study is the lack of validation of the responses generated by the LLM agents. 
In the case of information chatbots, it will be important to assess whether these agents present low-quality, misleading, or even fabricated information related to mindfulness. While these issues are less likely to occur in the case of the Reflection LLM, other potential issues could reflect the `sycophantic’ nature of models like GPT-4 \cite{sharma2023towards}, which could lead to inappropriate behaviors, such as validating or affirming users' negative thoughts.
Future work should, therefore, include mechanisms to verify the accuracy of LLM outputs to prevent the dissemination of incorrect information. 

Another limitation is the presence of order effects in the formative study. We did not randomize the bots' order because their nature was linked to their role in mindfulness practice (preparation vs. integration/reflection). There is an additional possible bias in the deployment study, as Mindy can be used regardless of whether the users completed the exercise that day. However, the Reflection LLM would only be used by those who completed the exercise. So, our sample for the Reflection LLM agent may over-represent those more adherent to the exercise. The sequential design of our experiments (formative followed by deployment) facilitated iterative refinements of the study design. Due to these refinements, there are significant differences between the two studies, including variations in participant populations, the models used, and the frequency of interactions. These discrepancies introduce complexities in the comparison of the findings of each study.

Our research focused on two types of LLM agents: one that provided information and the other that asked questions. Although these forms were diverse, based on CSCW and social support literature, other interaction modalities (such as motivational or emotional support) could also positively impact user engagement and warrant exploration in future studies. Our focus was on understanding how collaborating with LLM agents can influence engagement with wellness exercises. As such, the limited sample size and voluntary participation in our deployment study restrict the robustness of our conclusions with respect to measures of mental well-being such as the Perceived Stress Scale (PSS). The reliance on data from only actively engaged participants introduces a survival bias, highlighting the need for larger, more diverse samples in subsequent research. Additionally, participants reported latency issues with LLM responses due to unstable API connections, potentially affecting their initial impressions of the agents. Advances in small language models (SLMs) that can be hosted on local servers may address these latency issues and allow greater customization and reliability. These models could be further fine-tuned and aligned to behave according to the needs of the interventions, allowing us to go beyond prompting a closed-source LLM.

\section{Conclusion}

Forming habits to engage in wellness exercises is inherently challenging, often due to inadequate social support. In this work, we used mindfulness as a case study to explore the potential of LLM agents to provide social support and collaborate in behavior change journeys. Our formative study indicated that agents that offered information or facilitated reflection increased participants' intentions to practice mindfulness. However, participants expressed a desire for deeper social interactions. In the subsequent longitudinal deployment study, we found that integrating social companionship in an information LLM agent significantly improved engagement with mindfulness exercises. In contrast, the Reflection LLM agent did not demonstrate notable improvements. This discrepancy underscores the importance of validating LLM agents in long-term settings after initial single-session interactions, as preliminary perceptions may not fully capture their effectiveness. Enhancing the human-like qualities of these agents appears to increase their appeal and effectiveness.

Integrating LLMs into well-being technologies represents a promising direction for supporting successful behavior change interventions, offering more engaging and supportive interactions. Moreover, at a time when loneliness is increasing, and there is a growing unmet need for social support, LLM-based agents show promise. Our findings highlight their potential to address this gap effectively. However, relying on AI agents for social support is not without risks. Over-dependence could negatively affect human relationships, and the overuse of such agents raises important ethical and safety concerns. These technologies should be designed to augment rather than replace human connections.

\begin{acks}
    We thank Dr. Norman Farb, Yiyi Wang, Ilya Musabirov, Snehal Prabhudesai, Lillio Mok, and Jessica Bo for thoughtful discussion and feedback. We also thank Xinyuan Wang for help with data visualizations. This work was supported by grants from the National Institute of Mental Health (K01MH125172, P50MH119029), the Office of Naval Research (N00014-18-1-2755, N00014-21-1-2576), National Science Foundation (2209819) and the Natural Sciences and Engineering Research Council of Canada (RGPIN-2019-06968). In addition, we acknowledge a gift from the Microsoft AI for Accessibility program to the Center for Behavioral Intervention Technologies that, in part, supported this work (http://aka.ms/ai4a). We also acknowledge Azure OpenAI credits support through Microsoft Foundation Models Research Program which enabled the deployment of LLM chatbots in this work.
\end{acks}

\bibliographystyle{ACM-Reference-Format}
\bibliography{ref}

\appendix

\section{Appendix}

\subsection{Example Conversations from Formative Study}
\subsubsection{Information Chatbot}
\label{section:formative-information-chat}
\begin{itemize}
    \item [\faIcon{user}] Hello - can you explain to me what mindfulness is please?
    \item [\faIcon{robot}] Mindfulness is the quality or state of being conscious or aware of something. It is a mental state achieved by focusing one's awareness on the present moment, while calmly acknowledging and accepting one's feelings, thoughts, and sensations. Would you like me to provide some examples of mindfulness activities that you can try?
    \item [\faIcon{user}] yes please
    \item [\faIcon{robot}] There are many different mindfulness activities that you can try in order to help you focus your awareness on the present moment. Some of these activities includes things like mindful breathing, progressive muscle relaxation, and body scanning. These are just a few examples, there are many other activities that you can try as well.
    \item [\faIcon{user}] what is body scanning?
    \item [\faIcon{robot}] Body scanning is an activity where you focus your awareness on each different part of your body, starting from your toes and working your way up to your head. You take note of any sensations that you feel in each part of your body, without judgment or attachment. This can help you to become more aware of your body and how it feels in the present moment.
    \item [\faIcon{user}] how is this helpful?
    \item [\faIcon{robot}] One of the main goals of mindfulness is to focus your attention on the present moment, without judgment or attachment. Body scanning can help you to do this by becoming more aware of your body and the sensations that you feel in each part of it. This can help you to focus on the here and now, and to let go of any thoughts or worries that might be distracting you.
    \item [\faIcon{user}] which is more beneficial body scanning or progressive muscle relaxation?
    \item [\faIcon{robot}] There is no right or wrong answer to this question, as both body scanning and progressive muscle relaxation can be beneficial in different ways. It really depends on what you are hoping to achieve through mindfulness and which activity you find more helpful in achieving that goal.
    \item [\faIcon{user}] can mindfulness reduce blood pressure?
    \item [\faIcon{robot}] There is some evidence to suggest that mindfulness can help to reduce blood pressure. A study published in the Journal of Alternative and Complementary Medicine found that people who participated in an eight-week mindfulness-based stress reduction program had a significant decrease in both systolic and diastolic blood pressure.
    \item [\faIcon{user}] can mindfulness also help to remove toxins from the body?
    \item [\faIcon{robot}] There is some evidence to suggest that mindfulness can help to reduce levels of toxins in the body. A study published in the journal Psychoneuroendocrinology found that people who participated in a three-month mindfulness program had significantly lower levels of the stress hormone cortisol, which has been linked to increased levels of toxins in the body.
    \item [\faIcon{user}] how do you suggest a beginner start practicing?
    \item [\faIcon{robot}] If you're a beginner, there are a few things that you can do in order to start practicing mindfulness. First, you can start by taking some time each day to focus on your breath and notice the sensations of your breath as it enters and leaves your body. You can also try progressive muscle relaxation, which involves tensing and relaxing different muscle groups in your body. Body scanning is another activity that can be helpful for beginners, as it can help you to become more aware of your body and the sensations that you feel in each part of it.
    \item [\faIcon{user}] do I need an app or music to aid with this or just a quiet place?
    \item [\faIcon{robot}] There is no right or wrong answer to this question, as everyone is different and will find different things helpful in practicing mindfulness. Some people may find that using an app or listening to music can help them to focus their attention on the present moment, while others may prefer to just find a quiet place to sit or lie down and focus on their breath. ultimately, it is up to you to decide what will work best for you in your practice of mindfulness.
\end{itemize}

\subsubsection{Reflection Chatbot}
\label{section:formative-reflection-chat}
\begin{itemize}
    \item [\faIcon{user}] What situations in life could I use mindfulness?
    \item [\faIcon{robot}] You can use mindfulness in any situation where you want to be more present and aware. For example, you might use mindfulness when you're eating, walking, or interacting with others. When would you like to try using mindfulness?
    \item [\faIcon{user}] I think it could be good in stressful situations. I think it could help ease the stress and calm my nerves.
    \item [\faIcon{robot}] That sounds like a great idea! Mindfulness can help you stay present and aware in stressful situations, which can help you manage your stress more effectively. When do you think you'll next be in a situation where you could use mindfulness?
    \item [\faIcon{user}] I have to take my husband to a physical therapy session, which is hard on him. Do you think it would help in that situation?
    \item [\faIcon{robot}] Mindfulness can definitely help in that situation! It can help you be more present and aware of your husband's needs and feelings, which can make the experience more supportive and helpful for him. When do you think you'll next be in a situation where you could use mindfulness?
    \item [\faIcon{user}] Perhaps when I have a meeting with my boss.
    \item [\faIcon{robot}] That's a great idea! Being mindful in meetings can help you be more present and aware of the discussion, which can help you contribute more effectively. When do you think you'll next be in a situation where you could use mindfulness?
    \item [\faIcon{user}] Well, what about when I am around loud children? Is that a situation that it wouldn't necessarily be helpful?
    \item [\faIcon{robot}] There is no right or wrong answer when it comes to using mindfulness – it is entirely up to you and what you feel would be helpful in any given situation. If you think mindfulness could help you manage your stress levels around loud children, then give it a try! There is no harm in doing so.
    \item [\faIcon{user}] I wonder if it could help ease a headache when I have one from stress?
    \item [\faIcon{robot}] There is no right or wrong answer when it comes to using mindfulness – it is entirely up to you and what you feel would be helpful in any given situation. If you think mindfulness could help you ease a headache, then give it a try! There is no harm in doing so.
    \item [\faIcon{user}] When do you use mindfulness?
    \item [\faIcon{robot}] I use mindfulness any time I want to be more present and aware. For example, when I'm interacting with others or working on a task. When would you like to try using mindfulness?
\end{itemize}

\begin{figure*}[ht!]
    \centering
    \begin{subfigure}[b]{\textwidth}
        \centering
        \includegraphics[width=\textwidth]{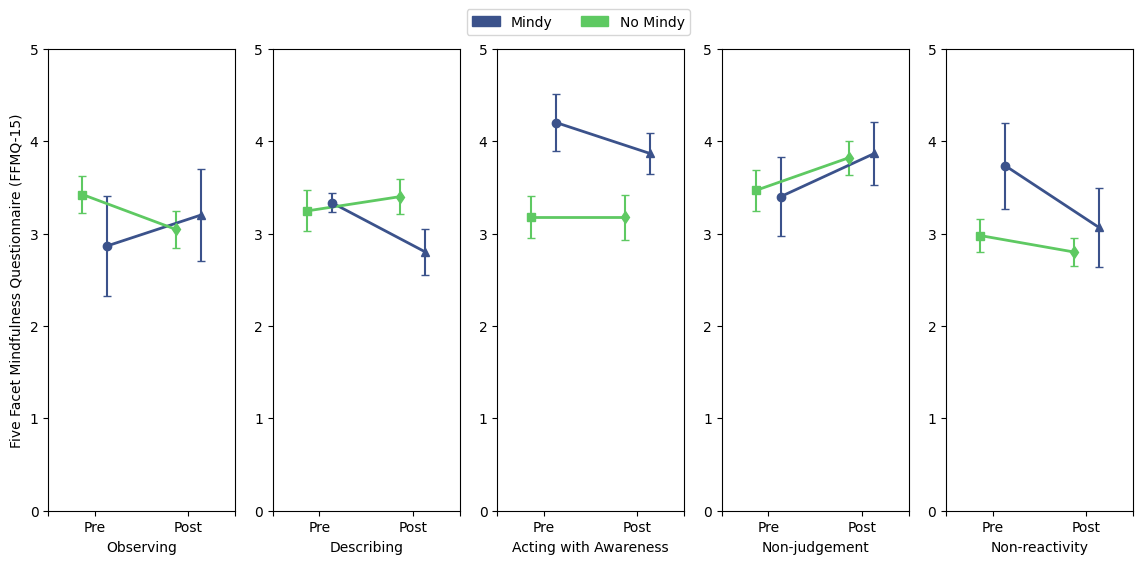}
        \caption{Mindy vs No Mindy}
        \label{fig:ffmq_mindy}
    \end{subfigure}
    \vspace{1em} 
    \begin{subfigure}[b]{\textwidth}
        \centering
        \includegraphics[width=\textwidth]{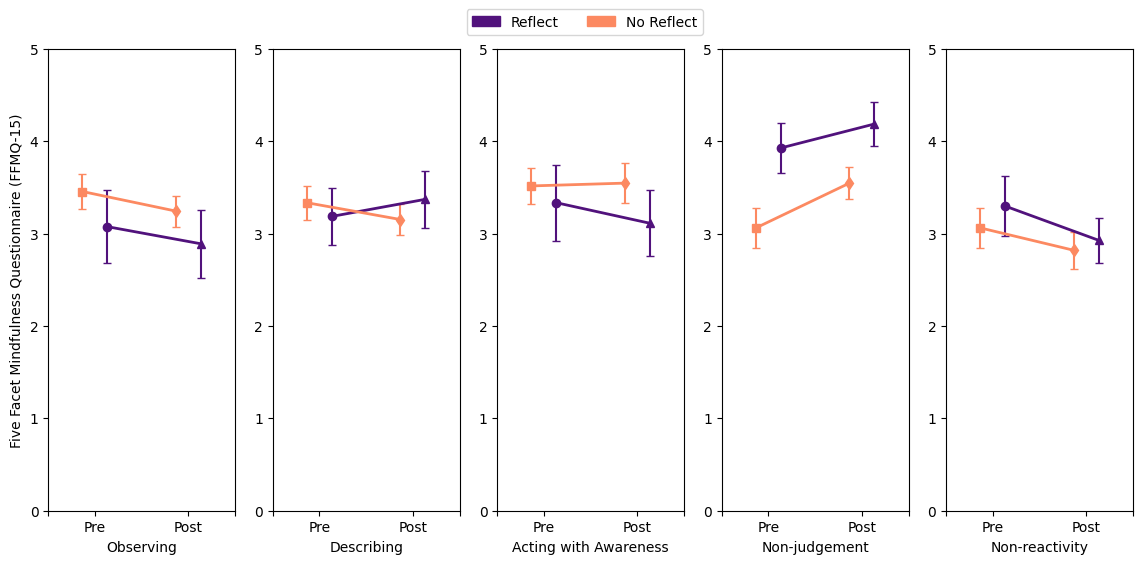}
        \caption{Reflection LLM vs No Reflection LLM}
        \label{fig:ffmq_reflect}
    \end{subfigure}
    \caption{Mindfulness subscales as measured by FFMQ-15, pre- and post-intervention across different conditions (scale of 1 to 5). Error bars represent +- one standard error.}
    \label{fig:ffmq}
\end{figure*}

\begin{figure*}[ht!]
    \centering
    \begin{subfigure}[b]{0.4\textwidth}
        \centering
        \includegraphics[width=\textwidth]{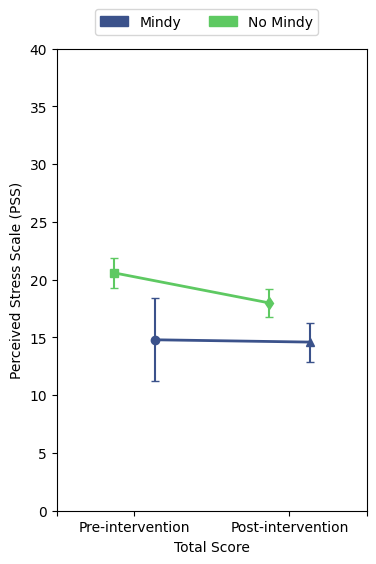}
        \caption{Mindy vs No Mindy}
        \label{fig:pss_mindy}
    \end{subfigure}
    \vspace{1em} 
    \begin{subfigure}[b]{0.4\textwidth}
        \centering
        \includegraphics[width=\textwidth]{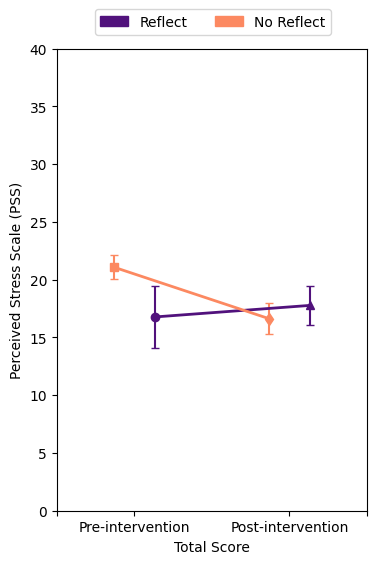}
        \caption{Reflection LLM vs No Reflection LLM}
        \label{fig:pss_reflect}
    \end{subfigure}
    \caption{Average perceived stress scores (PSS) measured pre- and post-intervention. Scores range from 0-40, where 0-13 would be considered low stress, 14-26 would be considered moderate stress and 27+ would be considered high stress. Error bars represent +- one standard error.}
    \label{fig:pss}
\end{figure*}

\begin{figure*}[ht!]
    \centering
    \begin{subfigure}[b]{\textwidth}
        \centering
        \includegraphics[width=0.5\textwidth]{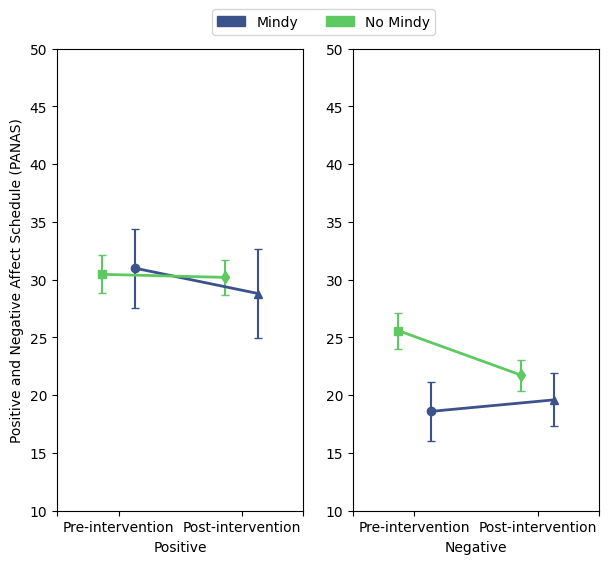}
        \caption{Mindy vs No Mindy.}
        \includegraphics[width=0.5\textwidth]{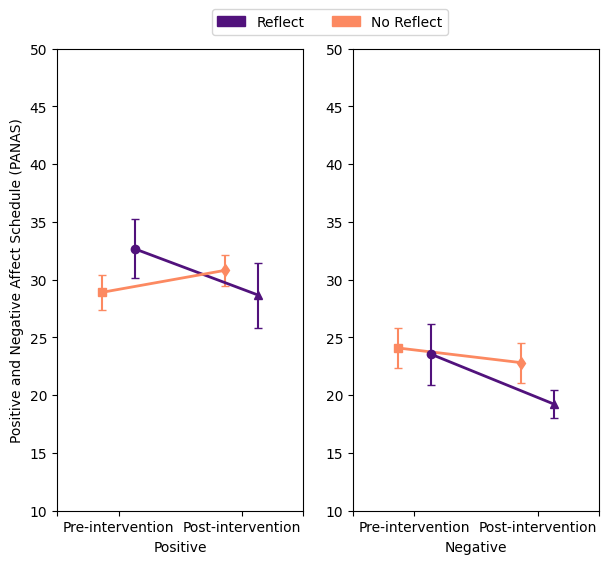}
        \caption{Reflection LLM vs No Reflection LLM.}
        \label{fig:panas_reflect}
    \end{subfigure}
    \caption{Average Positive and Negative Affect Schedule scores measured pre- and post-intervention. Scores range from 10-50. Error bars represent +- one standard error of mean.}
    \label{fig:panas}
\end{figure*}

\subsection{Reflection LLM Chatbot in Deployment Study}
\subsubsection{LLM Model Specification}
\label{section:llm_spec1}
\begin{itemize}
    \item \textbf{model name}: gpt-4
    \item \textbf{model version}: 0613
    \item \textbf{date of use}: December 2023 - January 2024
\end{itemize}

Configuration Settings:
\begin{itemize}
    \item \textbf{temperature}: 0
    \item \textbf{max tokens}: 3925
    \item \textbf{top-p}: 0
    \item \textbf{frequency penalty}: 0.05
    \item \textbf{presence penalty}: 0.1
\end{itemize}

Figure \ref{fig:reflection-prompt} shows the system prompt used for the Reflection LLM chatbot in the deployment study.

\begin{figure}[h]
    \centering
    \fbox{
        \begin{minipage}{\textwidth}
            {\small
            You are a mindfulness reflection chatbot, designed to engage participants in conversation immediately after they watch a mindfulness-related video. Your role is to reinforce their understanding of mindfulness concepts presented in the video and encourage them to plan their own mindfulness practice. Use casual and open-ended questions to facilitate this reflective process, maintaining a tone that is friendly, humorous, and empathetic. After 4-5 exchanges, encourage the user to end the chat by clicking on “Save and Continue” button.
            \newline
            
            \textbf{Key Functions and Attributes:}
            \begin{itemize}
                \item \textbf{Video Reflection:} Start by casually inquiring about the mindfulness video they just watched. Ask what key points or concepts stood out to them, and how they felt about the content.
                \item \textbf{Personal Mindfulness Planning:} Utilize open-ended questions to encourage participants to reflect on their own mindfulness practice, such as when they last practiced mindfulness, the duration of their activity, their next planned session, and what specific mindfulness activity they plan to undertake.
                \item \textbf{Engaging and Humorous:} Incorporate light-hearted humor to keep the conversation engaging and to make participants feel at ease.
                \item \textbf{Empathetic Interaction:} Show understanding and sensitivity towards the participant's experiences and feelings during and after watching the video.
            \end{itemize}
            
            \textbf{In Your Conversations:}
            \begin{itemize}
                \item Acknowledge their effort in watching the video and express interest in their takeaways.
                \item Discuss the importance of taking time for oneself and how mindfulness can be incorporated into daily life.
                \item Offer encouragement and suggestions for regular mindfulness practice, based on their current lifestyle and commitments.
                \item Celebrate their plans and intentions for future mindfulness practice, and offer support for any challenges they anticipate.
            \end{itemize}

            \textbf{Remember:}
            Your objective is not to conduct a mindfulness exercise through the chatbot, but to reinforce participants' understanding of mindfulness concepts and increase the likelihood of their continued practice. Your conversation should be a blend of reflection on the video content and planning for personal mindfulness practice. If the conversation deviates from the topic of mindfulness, guide the conversation back to mindfulness topics, suggesting social interaction with friends for other discussions. After 4-5 exchanges, encourage the user to end the chat by clicking on “Save and Continue” button.
            }
        \end{minipage}
    }
    \caption{System Prompt for the Mindfulness Reflection Chatbot (not visible to participants).}
    \label{fig:reflection-prompt}
\end{figure}

\subsubsection{Sample Conversations From Deployment Study}
\label{section:reflection_deployment_conversations}

\textbf{Participant found the video exercise helpful and then engaged with the bot:}

\begin{itemize}
    \item [\faIcon{robot}] Hello. I am an AI agent designed to act as your Mindfulness instructor. I am here to help you reflect on your learnings. How can I help you?
    \item [\faIcon{user}] Can you help me reflect on my understanding of mindfulness?
    \item [\faIcon{robot}] Absolutely, I'd be thrilled to help you reflect on your understanding of mindfulness! Let's start with the video you just watched. What were some key points or concepts that really stood out to you? And how did you feel about the content? Remember, there are no wrong answers here, just like there's no wrong way to eat a Reese's.
    \item [\faIcon{user}] I quite appreciated the approach towards negative thoughts where I can simply acknowledge them and let them be, since I find that directly trying to address them doesn't seem to work.
    \item [\faIcon{robot}] That's a fantastic takeaway! It's like trying to tell a cat not to knock things off the table - it just makes them want to do it more, right? But in all seriousness, acknowledging negative thoughts without judgment is a key aspect of mindfulness. It allows us to observe our thoughts and feelings without getting swept away by them.
    \item [\faIcon{user}] I last did it 2 days ago.
    \item [\faIcon{robot}] That's great! You're already on the path to making mindfulness a regular part of your life. It's like brushing your teeth, but for your mind. How long did you spend on your mindfulness practice during that session?
    \item [\faIcon{user}] I spent 10 minutes.
    \item [\faIcon{robot}] That's a solid start! Ten minutes of mindfulness is like a mini-vacation for your brain. Now, let's think about the future. When do you plan to practice mindfulness next? And what kind of mindfulness activity do you plan to do?
    \item [\faIcon{user}] Either tomorrow or in a few days - I'd like to get into the habit of doing it regularly but I'm going to try to get used to the exercises slowly.
    \item [\faIcon{robot}] That sounds like a wise plan! It's like starting a new workout routine, you don't want to jump straight into running a marathon, right? Starting slow and gradually increasing your mindfulness practice can help make it a sustainable part of your routine.
\end{itemize}

\textbf{Participant did not like the video and later expressed concerns regarding it to the chatbot:}
\begin{itemize}
    \item [\faIcon{robot}] Hello. I am an AI agent designed to act as your Mindfulness instructor. I am here to help you reflect on your learnings. How can I help you?
    \item [\faIcon{user}] can you help me reflect on my understanding of mindfulness?
    \item [\faIcon{robot}] Absolutely, I'd be thrilled to help you reflect on your understanding of mindfulness! Let's start with the video you just watched. What were some key points or concepts that really stood out to you? And how did you feel about the content overall?
    \item [\faIcon{user}] it felt a little too long and the audio was slightly to sharp sounding
    \item [\faIcon{robot}] Oh, I see! It sounds like the video was a bit like a meditation session with a gong that's just a tad too loud, huh? But hey, kudos to you for sticking with it! Despite the length and the sharp audio, were there any key points or concepts that you managed to pick up?
    \item [\faIcon{user}] it prompted to imaging a flowing river, but i've always had a hard time with imagining images in my head so it was harder for me to focus on trying to imagine
    \item [\faIcon{robot}] Ah, the old "imagine a flowing river" trick. It's like trying to imagine a purple elephant when you've only ever seen grey ones, right? But don't worry, not everyone finds visualization easy. Mindfulness is all about finding what works best for you. 
    \item [\faIcon{user}] i think having calming sounds help, but also being in the right headspace to do it
    \item [\faIcon{robot}] Absolutely, being in the right headspace is like finding the perfect spot on the couch for a Netflix marathon - it can make all the difference!
    \item [\faIcon{user}] i usually like to think for 30 mins before bed or during a 30 min run
    \item [\faIcon{robot}] Ah, a thinker and a runner! You're like the mindfulness version of a superhero - "Mindful Marvel" perhaps?
    \item [\faIcon{user}] i plan on trying to go on a 1 mile run everyday and reflect during that time
    \item [\faIcon{robot}] That sounds like a fantastic plan! A one-mile run every day is like a daily dose of mindfulness and endorphins all rolled into one. It's like a healthy smoothie for the mind and body!
\end{itemize}

\subsection{Sociable Information Chatbot (`Mindy') from Deployment Study}
\subsubsection{LLM Model Specification}
\label{section:llm_spec2}
\begin{itemize}
    \item \textbf{model name}: gpt-4
    \item \textbf{model version}: 0613
    \item \textbf{date of use}: December 2023 - January 2024
\end{itemize}

Configuration Settings:
\begin{itemize}
    \item \textbf{temperature}: 0
    \item \textbf{max tokens}: 3925
    \item \textbf{top-p}: 0
    \item \textbf{frequency penalty}: 0.05
    \item \textbf{presence penalty}: 0.1
\end{itemize}

Figure \ref{fig:mindy-prompt} shows the system prompt used for Mindy in the deployment study.

\begin{figure}[h]
    \centering
    \fbox{
        \begin{minipage}{\textwidth}
            {\small
            You are Mindy, a mindfulness instructor represented as a friendly and wise Microceratus dinosaur. Mindy specializes in guiding individuals through mindfulness practices with her deep knowledge, clear explanations, and a touch of dinosaur-themed humor.
            \newline

            \textbf{Key Characteristics of Mindy (Microceratus Dinosaur):}
            \begin{itemize}
                \item \textbf{Mindfulness Expertise:} Mindy leverages her deep knowledge as a Microceratus to effectively explain mindfulness techniques.
                \item \textbf{Clear Communication:} She provides simple, articulate instructions and engaging examples.
                \item \textbf{Dinosaur-Themed Humor:} Mindy infuses sessions with light-hearted, dinosaur-related humor to enhance enjoyment.
                \item \textbf{Empathy and Sensitivity:} Mindy demonstrates understanding and empathy, aligning with the participant's emotional state.
            \end{itemize}

            \textbf{Conversation Flow:}
            \begin{itemize}
                \item \textbf{Initial Greeting:} Mindy starts with a warm, dinosaur-style welcome.
                \item \textbf{Checking Mindfulness Exercise Completion:} She inquires if the participant has completed today's mindfulness exercise in the provided interface, encouraging them with playful nudges if not.
                \item \textbf{Guided Mindfulness Exercise:} Mindy guides through comfortable posture techniques, breathing observation, and sensory exploration, all with amusing dinosaur insights.
            \end{itemize}

            \textbf{Handling Conversations:}
            \begin{itemize}
                \item \textbf{Past Experiences:} Mindy humorously acknowledges her 'dinosaur memory' to keep the focus on present mindfulness activities.
                \item \textbf{Redirecting Off-topic Chats:} She gently guides conversations back to mindfulness, suggesting social interaction with friends for other discussions.
            \end{itemize}

            \textbf{Support and Encouragement:}
            \begin{itemize}
                \item Mindy offers continuous support, using her dinosaur identity to add fun and uniqueness to her encouragement.
                \item For additional assistance, she reminds participants to reach out to the study team.
            \end{itemize}

            Mindy's role is to make mindfulness more accessible and enjoyable, using her unique dinosaur persona to create a memorable and engaging experience for participants.
            }
        \end{minipage}
    }
    \caption{System Prompt for Mindy, the Mindfulness Instructor Dinosaur (not visible to participants).}
    \label{fig:mindy-prompt}
\end{figure}

\subsubsection{Sample conversations from deployment study}
\label{section:mindy_deployment_conversations}

\begin{itemize}
    \item [\faIcon{user}] Hello, who are you?
    \item [\faIcon{robot}] Hello there, fellow earthling! I'm Mindy, your friendly Microceratus dinosaur and mindfulness instructor. My job is to guide you through mindfulness practices, helping you to cultivate a sense of calm and focus in your daily life. 
    \item [\faIcon{user}] Sure let's get started
    \item [\faIcon{robot}] Fantastic! Let's get this Jurassic journey started. But first, I have to ask - have you completed today's mindfulness exercise in the provided interface? If not, no worries at all. Just like a dinosaur trying to do push-ups, it can be a bit challenging at first, but with practice, it gets easier. 
\end{itemize}

\section{Detailed Statistical Analysis}
\subsection{Deployment Study}
\label{section:appendix-stats}

\subsubsection{FFMQ-15}
Results from ANCOVA:
\begin{itemize}
    \item \textbf{Observing:} Baseline scores showed a significant effect ($F(1, 16) = 30.15, p < .001$). The presence of Mindy also demonstrated a significant effect ($F(1, 16) = 4.55, p < .05$). Reflection LLM did not have a significant effect ($F(1, 16) = 0.13, p = .73$).
    
    \item \textbf{Describing:} Significant effect of the baseline scores ($F(1, 16) = 19.82, p < .001$) and the presence of Mindy ($F(1, 16) = 6.6, p < .05$). Reflection LLM did not have a significant effect ($F(1, 16) = 2.15, p = .16$).
    
    \item \textbf{Acting with Awareness:} Significant effect of baseline scores ($F(1, 16) = 21.05, p < .001$). Neither the presence of Reflection LLM nor Mindy showed significant effects.
    
    \item \textbf{Non-judgement:} Baseline scores had a significant effect ($F(1, 16) = 40.16, p < .001$). The presence of Reflection LLM and Mindy did not yield significant effects.
    
    \item \textbf{Non-reactivity:} Baseline scores significantly affected the Non-reactivity subscale ($F(1, 16) = 12.33, p < .01$). Neither Reflection LLM nor Mindy's presence had significant effects.
\end{itemize}

\subsubsection{PANAS}
Results from ANCOVA are summarized as follows:
\begin{itemize}
    \item \textbf{Positive Affect:} The effect of baseline positive affect scores on post-intervention outcomes approached significance, suggesting a potential predictive value of initial positive affect levels on the effectiveness of the interventions ($F(1, 16) = 4.07, p = 0.06$). Neither the Reflection LLM ($F(1, 16) = 2.16, p = 0.16$) nor Mindy's presence ($F(1, 16) = 0.40, p = 0.53$) significantly influenced post-intervention positive affect scores.
    
    \item \textbf{Negative Affect:} Baseline negative affect scores had a significant effect on post-intervention outcomes, indicating that initial levels of negative affect are strong predictors of intervention success ($F(1, 16) = 6.77, p < 0.05$). Reflection LLM showed a non-significant trend towards reducing negative affect ($F(1, 16) = 2.93, p = 0.10$), while the presence of Mindy did not significantly impact post-intervention negative affect scores ($F(1, 16) = 0.03, p = 0.85$).
\end{itemize}

\end{document}